\documentclass[a4paper,11pt]{article}
\pdfoutput=1 

\usepackage{jheppub} 

\usepackage{natbib}
\usepackage[T1]{fontenc} 

\usepackage{xspace}
\usepackage{graphicx}
\usepackage{multicol}
\usepackage{lineno}
\usepackage{amsmath}
\usepackage{subcaption}
\usepackage{physics}

\newcommand*\patchAmsMathEnvironmentForLineno[1]{%
  \expandafter\let\csname old#1\expandafter\endcsname\csname #1\endcsname
  \expandafter\let\csname oldend#1\expandafter\endcsname\csname end#1\endcsname
  \renewenvironment{#1}%
     {\linenomath\csname old#1\endcsname}%
     {\csname oldend#1\endcsname\endlinenomath}}%
\newcommand*\patchBothAmsMathEnvironmentsForLineno[1]{%
  \patchAmsMathEnvironmentForLineno{#1}%
  \patchAmsMathEnvironmentForLineno{#1*}}%
\AtBeginDocument{%
\patchBothAmsMathEnvironmentsForLineno{equation}%
\patchBothAmsMathEnvironmentsForLineno{align}%
\patchBothAmsMathEnvironmentsForLineno{flalign}%
\patchBothAmsMathEnvironmentsForLineno{alignat}%
\patchBothAmsMathEnvironmentsForLineno{gather}%
\patchBothAmsMathEnvironmentsForLineno{multline}%
}

\usepackage[output-decimal-marker={.}, separate-uncertainty=true]{siunitx}
\usepackage[section]{placeins}
\usepackage[colorlinks=true]{hyperref}
\usepackage[capitalise, noabbrev]{cleveref}

\crefname{equation}{eq.}{eqs.}
\crefname{figure}{figure}{figure}
\crefname{section}{section}{sections}
\Crefname{equation}{Eq.}{Eqs.}
\Crefname{figure}{Figure}{Figure}
\Crefname{section}{Section}{Sections}

\graphicspath{{_fig/}}

\newcommand{\ub}{MicroBooNE\xspace}

\newcommand{\numu}{$\nu_{\mu}$\xspace}
\newcommand{\numucc}{$\nu_{\mu} CC$\xspace}

\newcommand{\dedx}{$\differential E / \differential x$\xspace}

\newcommand{\pid}{$\mathcal{P}$\xspace}

\newcommand{\pitch}{local pitch\xspace}

\title{Calorimetric classification of track-like signatures in liquid argon TPCs using MicroBooNE data}
\collaboration{MicroBooNE Collaboration}
\author[jj]{P.~Abratenko}
\author[o]{R.~An}
\author[d]{J.~Anthony}
\author[ii]{J.~Asaadi}
\author[t,gg]{A.~Ashkenazi}
\author[mm,l]{S.~Balasubramanian}
\author[l]{B.~Baller}
\author[u]{C.~Barnes}
\author[y]{G.~Barr}
\author[s]{V.~Basque}
% ...

% ...
\affiliation[d]{University of Cambridge, Cambridge CB3 0HE, United Kingdom}
% ...
\affiliation[l]{Fermi National Accelerator Laboratory (FNAL), Batavia, IL 60510, USA}
% ...
\affiliation[o]{Illinois Institute of Technology (IIT), Chicago, IL 60616, USA}
% ...
\affiliation[s]{The University of Manchester, Manchester M13 9PL, United Kingdom}
\affiliation[t]{Massachusetts Institute of Technology (MIT), Cambridge, MA, 02139, USA}
\affiliation[u]{University of Michigan, Ann Arbor, MI, 48109, USA}
% ...
\affiliation[y]{University of Oxford, Oxford OX1 3RH, United Kingdom}
% ...
\affiliation[gg]{Tel Aviv University, Tel Aviv, Israel, 69978}
% ...
\affiliation[ii]{University of Texas, Arlington, TX, 76019, USA}
\affiliation[jj]{Tufts University, Medford, MA, 02155, USA}
% ...
\affiliation[mm]{Wright Laboratory, Department of Physics, Yale University, New Haven, CT, 06520, USA}

  \emailAdd{microboone\_info@fnal.gov}

\abstract{
The MicroBooNE liquid argon time projection chamber located at Fermilab is a neutrino experiment dedicated to the study of short-baseline oscillations, the measurements of neutrino cross sections in liquid argon, and to the research and development of this novel detector technology.
Accurate and precise measurements of calorimetry are essential to the event reconstruction and are achieved by leveraging the TPC to measure deposited energy per unit length along the particle trajectory, with mm resolution.
We describe the non-uniform calorimetric reconstruction performance in the detector, showing dependence on the angle of the particle trajectory.
Such non-uniform reconstruction directly affects the performance of the particle identification algorithms which infer particle type from calorimetric measurements.
This work presents a new particle identification method which accounts for and effectively addresses such non-uniformity.
The newly developed method shows improved performance compared to previous algorithms, illustrated by a 93.7\% proton selection efficiency and a 10\% muon mis-identification rate, with a fairly loose selection of tracks performed on beam data.
The performance is further demonstrated by identifying exclusive final states in \numucc interactions.
While developed using MicroBooNE data and simulation, this method is easily applicable to future LArTPC experiments, such as SBND, ICARUS, and DUNE.
}

\begin{document} 
\maketitle
\flushbottom

\section{Introduction}
\label{sec:intro}
\begin{figure}
\centering
\subfloat{\includegraphics[width=0.45\textwidth]{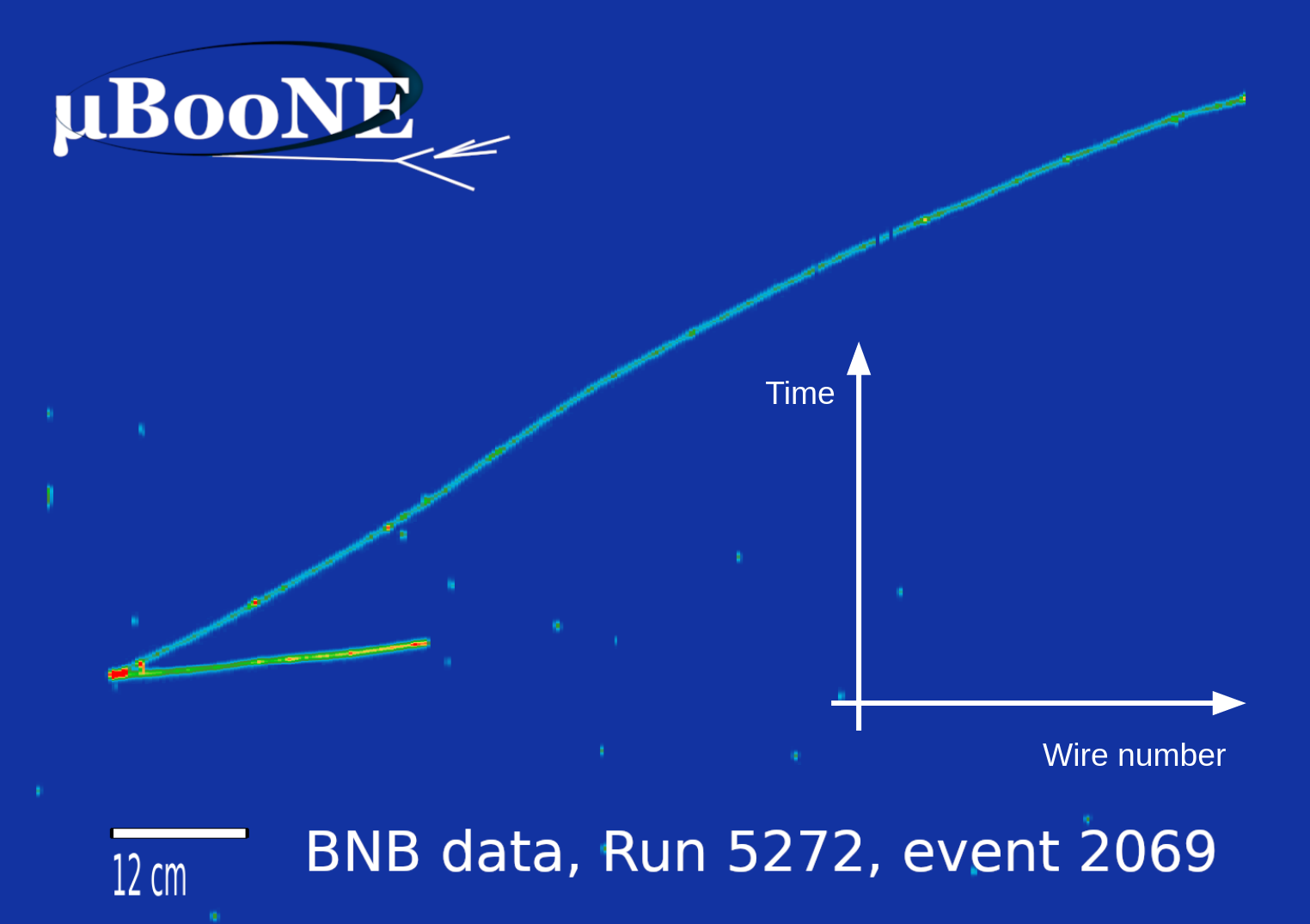}}\hfill
\subfloat{\includegraphics[width=0.55\textwidth]{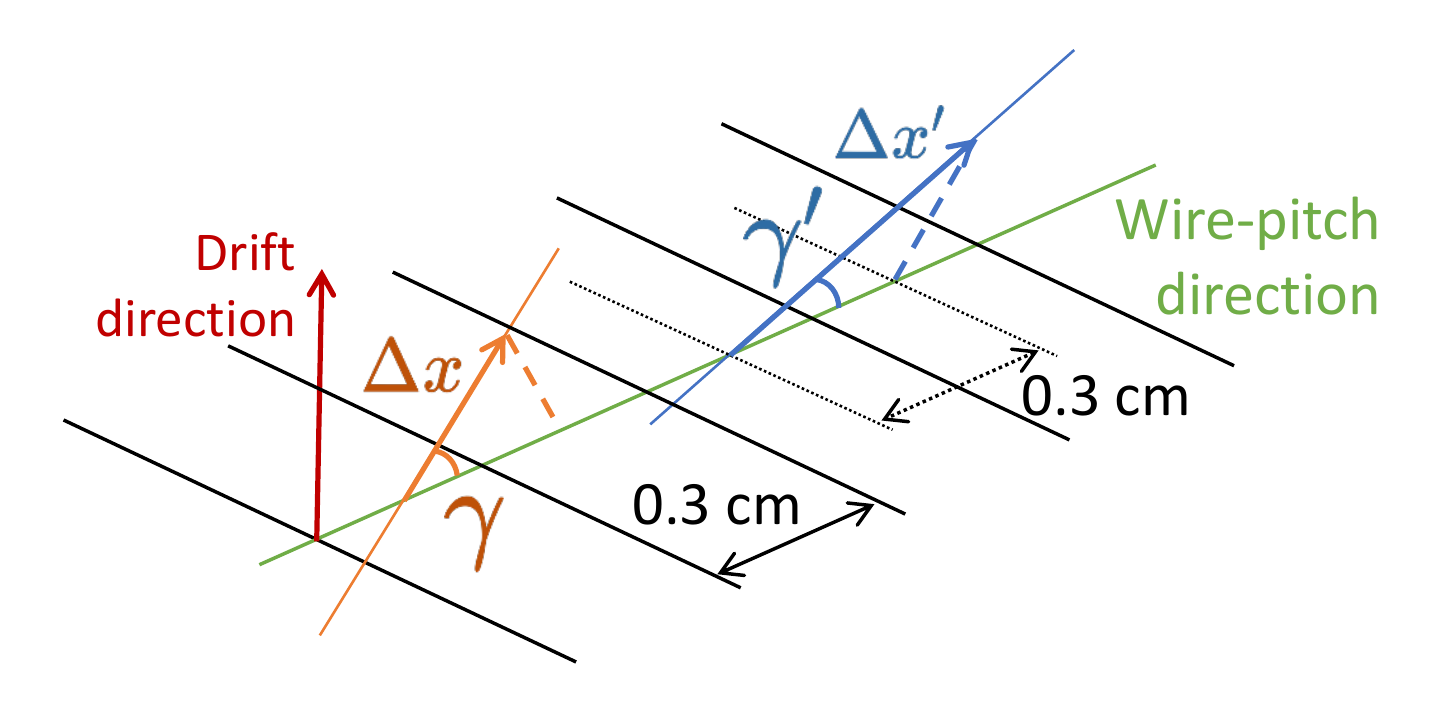}}
\caption{
Left: Example of a raw image of a \numucc interaction in \ub with two tracks in the final state, as recorded on the collection plane.
The deposited charge (color scale) is shown as a function of wire number (x-axis) and time (y-axis).
The ionization profiles of the two particles are used to identify them as one proton and one muon.
Right: A sketch of the relevant directions and angle in the calorimetric reconstruction.
The orange and blue arrows represent two possible particle trajectories with different gamma angles and different \pitch values, represented by their lengths.
Black solid lines represent wires, spaced 0.3 cm apart.
The dashed lines are perpendicular to the wire-pitch direction, and make evident the connection between the angle $\gamma$ and $\Delta x$, through \cref{eqn:pitch}.
}
\label{fig:intro_plot}
\end{figure}

Liquid argon time projection chambers (LArTPCs) are powerful neutrino detectors which enable the study of topological and calorimetric signatures of particles produced in neutrino interactions with millimeter spatial resolution \cite{icarus, argoneut, lariat}. 
These particles are reconstructed as track-like (or simply tracks) if they deposit energy predominantly through ionization, or shower-like (or simply showers), if the main energy loss mechanisms are bremsstrahlung and pair production. 
Tracks are primarily associated with the reconstruction of muons, protons, and pions, while showers are associated with electrons and photons.
At typical \ub energies, between hundreds of MeV to few GeV, hadrons do not produce hadronic showers.
However, thanks to the high resolution, if a LArTPC like \ub were to be operated at larger energies, the different particles produced in hadronic showers could be reconstructed individually.
Energy lost by final state charged particles results in the ionization of argon atoms. 
Trails of ionization electrons are detected by multiple wire planes to provide a 3D image of the particles' propagation through the detector. 
Particle identification (PID), determination of the type of particle given its calorimetric measurement, is performed by studying the profiles of each ionization electron trail.
The left panel of \cref{fig:intro_plot} shows the collection plane projection of a muon neutrino charged current (\numucc) interaction, in which two track-like particles are produced.
These tracks are classified as one proton and one muon, the two most common track-like particles in \ub, differentiated by the different amounts of energy deposited per unit length at any given point in their trajectories.

\begin{figure}
\centering
\subfloat{\includegraphics[width=0.98\textwidth]{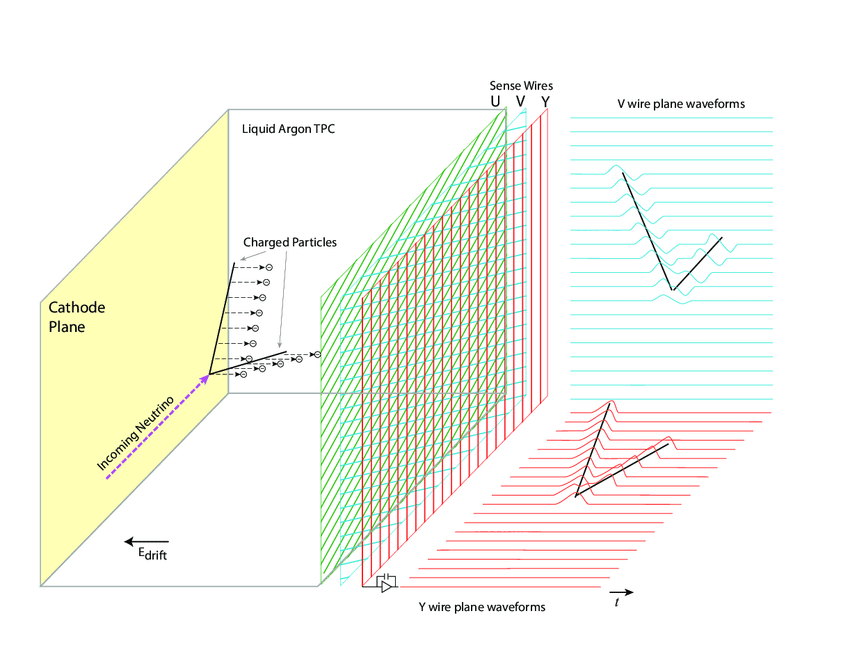}}\hfill
\caption{
In this schematic representation of the \ub LArTPC (see text for details), a neutrino interacts in the detector, producing two charged particles in the final state.
The ionization is drifted towards the anode, and induces waveforms on the wire planes, which are displayed for the V and Y planes.
Figure taken from \cite{microboone_experiment}.
}
\label{fig:detector_image}
\end{figure}

After a brief discussion of the \ub detector and reconstruction in \cref{sec:uboone}, in \cref{sec:angular_effects} we illustrate the angular dependence of LArTPCs' calorimetric reconstruction.
In \cref{sec:pid}, a review of the particle identification principles is presented.
In \cref{sec:likelihood} we address the central topic of the paper: a new method to perform particle identification that more easily accounts for angular dependencies in calorimetric reconstruction.
The performance of this method is discussed in \cref{sec:applications}, where the identification of different muon neutrino interaction final states is presented.

\section{The \ub detector and the calorimetric reconstruction of tracks}
\label{sec:uboone}
The MicroBooNE detector is an 85 ton active mass liquid argon time projection chamber (LArTPC) \cite{microboone_experiment}.
The drift direction ($x$), vertical direction ($y$), and the beam direction ($z$) measure 2.56\,m, 2.33\,m, and 10.36\,m, respectively.
The nominal electric field inside the TPC is 273.9\,V/cm resulting in an electron drift velocity of 0.11\,cm/$\mu$s. 
The drifted ionization charge from particle interactions is read out by three wire planes spaced 0.3\,cm apart, with a 0.3\,cm wire spacing, oriented vertically for the collection plane (Y plane), at +60 deg for the first induction plane (U plane), and at -60 deg for the second induction plane (V plane).
The digitized charge on the readout wires is noise-filtered \cite{noise_filtering} and deconvolved \cite{ub_signal_processing1, ub_signal_processing2}.
The high signal-to-noise ratio of \ub's TPC and cold electronics allow for the accurate measurement of charge on all three wire-planes, an essential ingredient in reliable particle identification.
The data is represented as a set of three two-dimensional images (one for each wire plane), with wire number plotted along the horizontal axis and drift time plotted along the vertical axis.
Each provides a two-dimensional projection of the charge deposited in the event.
An example is given in the left panel of \cref{fig:intro_plot}.

The position and amount of charge deposited is characterized by the hit-finding process: charge depositions on a given wire at a given time, called hits, are extracted through a Gaussian fit to the waveforms \cite{hit_finder}.
Next, the Pandora multi-algorithm pattern recognition framework \cite{pandora} groups nearby hits into clusters.
Clusters on the three planes are subsequently matched to reconstruct three dimensional particles.
Each cluster is a projection on a given plane of the charge deposited by a particle.
The deposited charge is corrected for detector effects to provide a spatially uniform response.
For each hit in a cluster the charge deposited is converted to deposited energy, providing a local measurement of the energy $\Delta E$ along a three-dimensional distance $\Delta x$ extracted from the reconstructed trajectory \cite{ub_calibration}.
The conversion is performed with a multiplicative factor specific to the plane, and correcting for the recombination of electron and ions \cite{ub_calibration, recombination_argoneut}.
The three-dimensional distance $\Delta x$ is called \textit{\pitch}, and is computed as 
\begin{equation}
\label{eqn:pitch}
    \Delta x = 0.3\,\text{cm}\,/ \cos (\gamma),
\end{equation}
where 0.3 cm is the wire spacing and $\gamma$ is the three-dimensional angle between the local direction of the track and the vector that connects adjacent wires (also called wire-pitch direction), as illustrated in the right panel of \cref{fig:intro_plot}.
The angle $\gamma$ ranges between 0 and 90 degrees, while $\Delta x$ takes values between 0.3 cm and infinity.
Combined with the measurement of $\Delta E$, the ionization density \dedx can be estimated for every hit.

\section{Angular effects in calorimetric reconstruction}
\label{sec:angular_effects}
When measuring calorimetric information in a LArTPC, the fact that charge is drifted along a particular direction (drift direction) and projected on wire planes with different orientations makes the calorimetric reconstruction angle-dependent.
Both \dedx and the precision with which it is measured depend on the direction of the ionization trace left by the particle, even in a "perfect detector", absent of detector effects and angle-dependent detector response non-uniformity.
The dependence appears primarily through the angle $\gamma$ illustrated in the bottom plot of \cref{fig:intro_plot}.
Because the angle $\gamma$ relates directly to the \pitch through a bijective function (\cref{eqn:pitch}), $\gamma$ and \pitch can be used interchangeably, and \pitch will be used in the rest of the article.

Measured \dedx, even with a perfect detector, is angle-dependent because of intrinsic statistical fluctuations in particle energy loss, which impact the probability density function of measurements when averaged over different travel distances (different \pitch).
A \dedx distribution, typically described by a Landau function for small \pitch values \cite{Landau_energy_loss}, becomes narrower at larger \pitch and its most probable value moves to higher \dedx values while its average remains constant.
This general geometrical effect applies to all LArTPCs and it is shown in figure 33.8 in \cite{pdg}.

The precision of \dedx measurements also depends on the \pitch since the shape of the signals induced on the wires by the drifting charge appears very different at lower \pitch (< 0.7 cm) compared to larger \pitch (> 0.7 cm) \cite{ub_signal_processing1}, impacting hit reconstruction and making measurements more precise at lower \pitch.
\begin{figure}[!ht]
\centering
\subfloat{\includegraphics[width=0.5\textwidth]{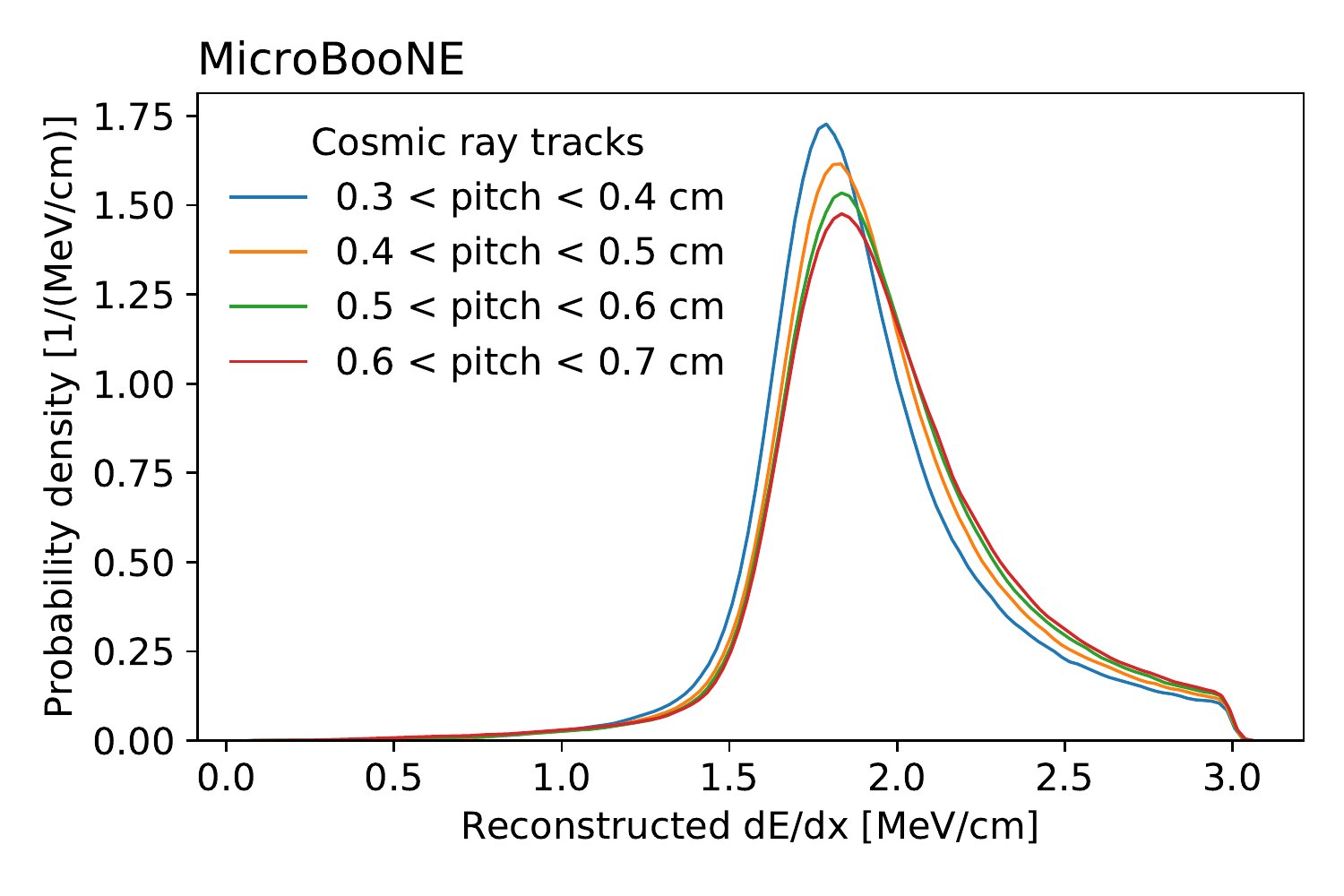}}\hfill
\subfloat{\includegraphics[width=0.5\textwidth]{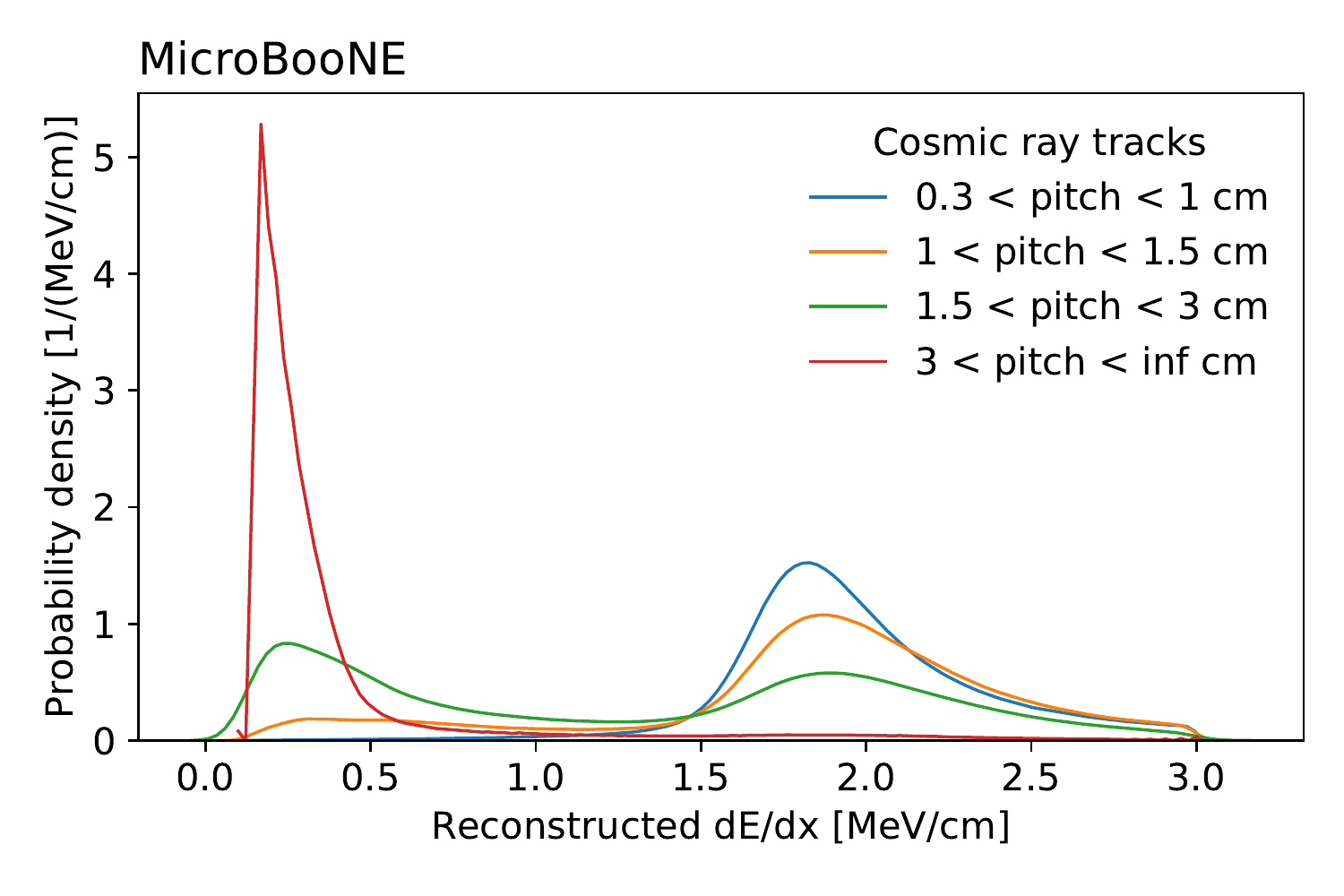}}
\caption{Normalized distributions of \dedx for different \pitch in the low-\pitch regime (left) and over the entire \pitch spectrum (right), as measured on the collection plane in a sample of cosmic muons tracks in the \ub data.}
\label{fig:dedx_vs_pitch}
\end{figure}
\Cref{fig:dedx_vs_pitch} illustrates these effects as measured with cosmic ray tracks in \ub's data.
By requiring the tracks to cross the detector, this selection results in a clean set of relativistic muons, which are minimum ionizing particles, with a constant average \dedx.
The left plot shows the shape of the \dedx distribution for different small \pitch values, where the peak of the distributions shifts towards larger \dedx values for larger \pitch, varying by 4\% in the low \pitch range (between 0.3 cm and 0.7 cm).
The width of the distributions increases at larger \pitch because the finite precision introduced by the detector smears out the distribution more than the predicted shrinking induced by geometrical effects. 
The right plot shows analogous distributions integrated over a wider range of \pitch values, illustrating a major change in the shape of the distributions at very large \pitch.
The distributions for larger \pitch values show a second peak at smaller \dedx values, mainly populated by particles traveling parallel to the drift direction.
These tracks induce long signals in time, for which the Gaussian fit performed by the hit-finding process is not sufficient.
Therefore, such signals are fit by a sum of Gaussian shapes, for which the overall number of hits, their positions, and their amplitudes are free parameters.
The width is fixed to reduce the degeneracy of the problem.
The total deposited charge is therefore segmented into multiple hits, leading to an underestimation of the hit charge and resulting in smaller \dedx values.
These small and non-physical values of \dedx encountered at large \pitch are not correlated with the true \dedx, bringing no additional information and reducing the particle identification performance.

Only considering the dependence on the angle $\gamma$, which is the polar angle with respect to the wire-pitch direction, is an approximation, as it encapsulates most, but not all the angular dependence.
A more complete analysis will consider the additional dependence on the relative azimuthal angle, that, together with $\gamma$, uniquely describes the 3D trajectory.
Nonetheless, this approximation captures most of the angular dependence, significantly improving particle identification performance.

\section{Energy deposition profile and particle identification}
\label{sec:pid}
Typical particle identification methods condense calorimetric information into a score used to distinguish different particle species.
The score is typically obtained by starting from the measured \dedx profile as a function of residual range - the distance of a given energy deposition within the track from the endpoint of the track itself.
This profile is compared with the expectation for different particle hypotheses to choose the hypothesis that best matches the data.
The hypothesized \dedx profile is computed by integrating the Bethe-Bloch function for a given particle mass and charge.
In performing this comparison, the intrinsic statistical nature of \dedx must be accounted for, as well as the angular dependencies (described in the previous section) which also affect the shape of the \dedx distribution.
As \pitch values are different on different readout planes, combining the three wire plane measurements ensures that the calorimetric information provided by the LArTPC in the entire $4\pi$ solid angle is fully leveraged for the purpose of performing particle identification.
\section{The likelihood-based method for particle identification}
\label{sec:likelihood}
The newly-developed method for particle identification accounts for angular dependencies in calorimetric reconstruction and compensates for distortions at large \pitch by efficiently combining the three wire plane measurements.

This method computes the likelihood for different particle hypotheses given the experimentally measured \dedx profile.
The likelihood is based on a model which is computed through an accurate description of the expected \dedx probability density function.

\subsection{The \texorpdfstring{\dedx}{dedx} probability density function}
\label{sec:pdf}
The \dedx probability density function (PDF) for each particle type is the basic ingredient of the likelihood calculation.
In principle, the average \dedx as a function of the residual range could be estimated by integrating the Bethe-Bloch function before applying detector reconstruction.
However, a complete characterization of the \dedx distribution requires an analytic description of the intrinsic fluctuations of the ionization energy loss and effect of the detector reconstruction. 
This is challenging given the very long computational time required and because there is no straightforward way to derive an analytic description of the detector reconstruction.
The \dedx PDF is instead estimated from the \ub simulation which incorporates all the described effects, as demonstrated in the data/simulation comparisons in \cref{sec:applications}.

\begin{figure}[!ht]
\centering
\subfloat {\includegraphics[width=0.33\textwidth]{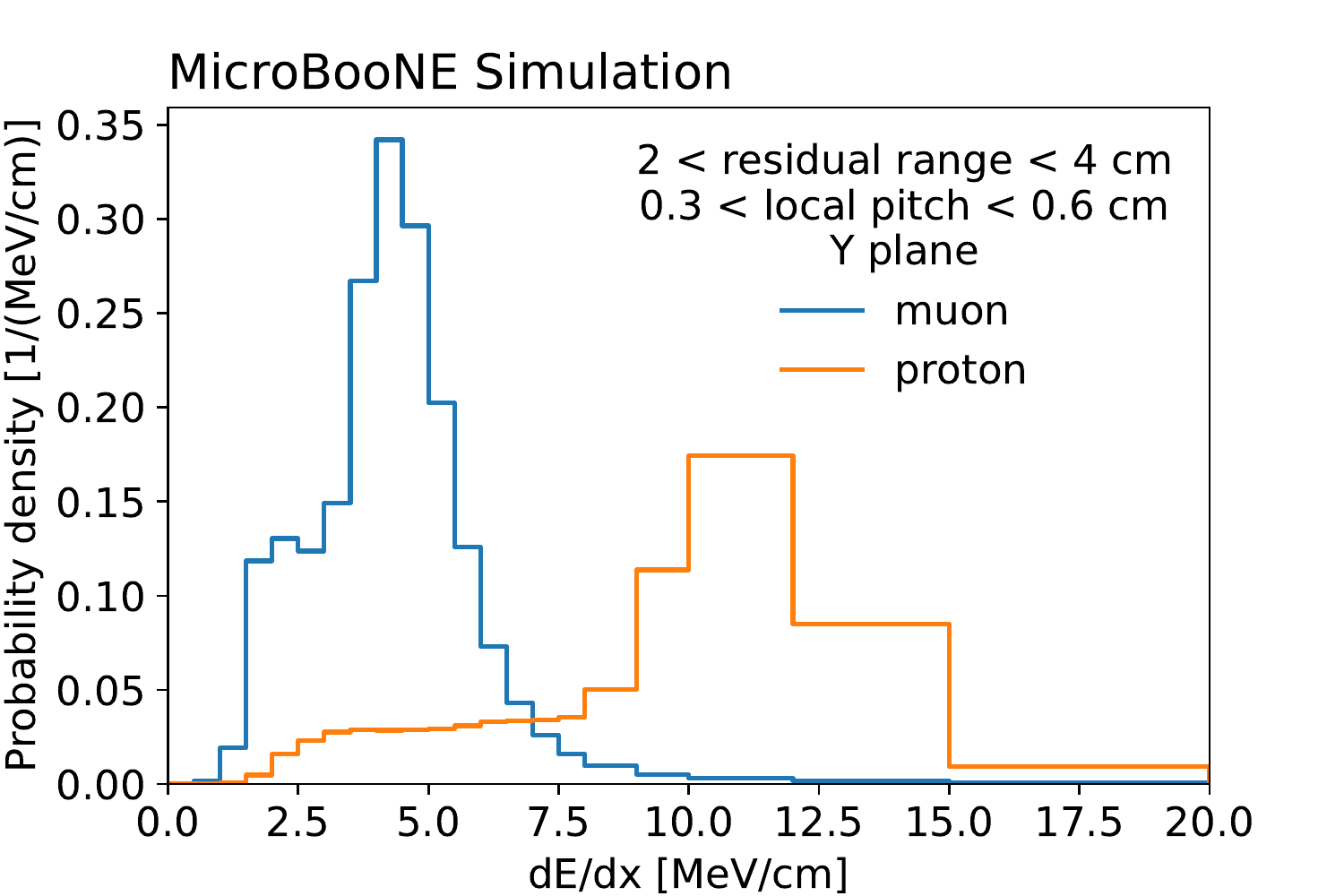}}
\subfloat {\includegraphics[width=0.33\textwidth]{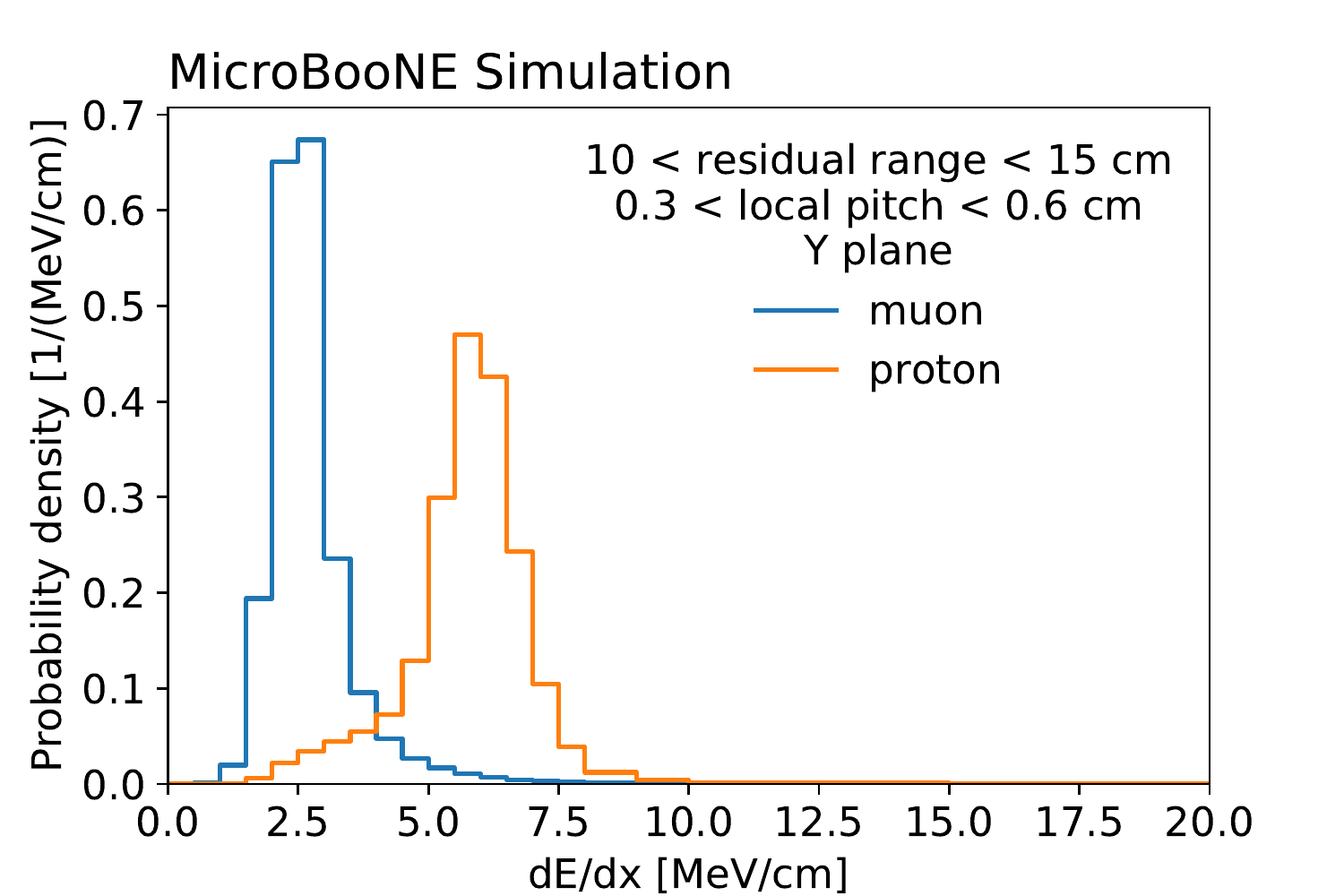}}
\subfloat {\includegraphics[width=0.33\textwidth]{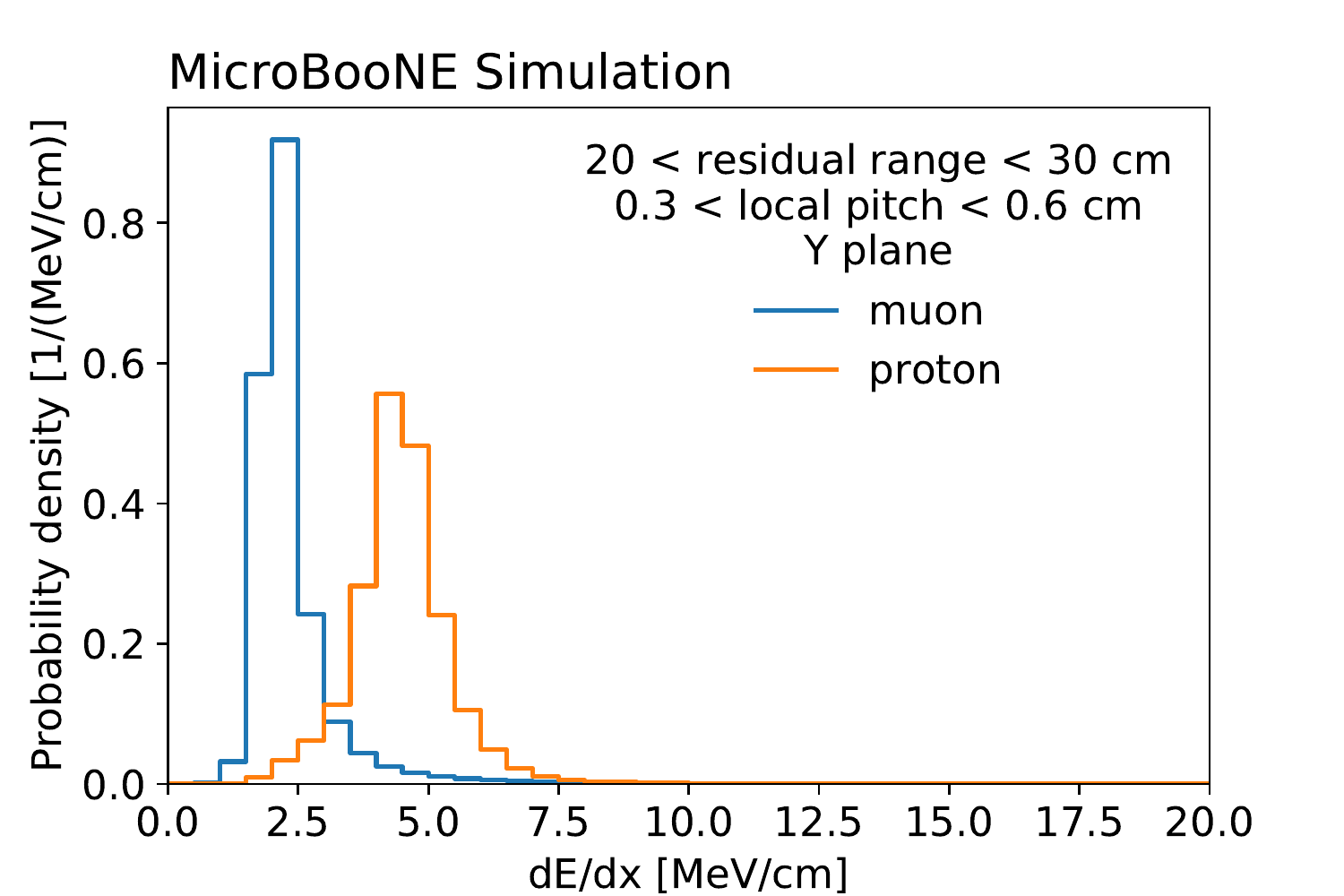}}

\subfloat {\includegraphics[width=0.33\textwidth]{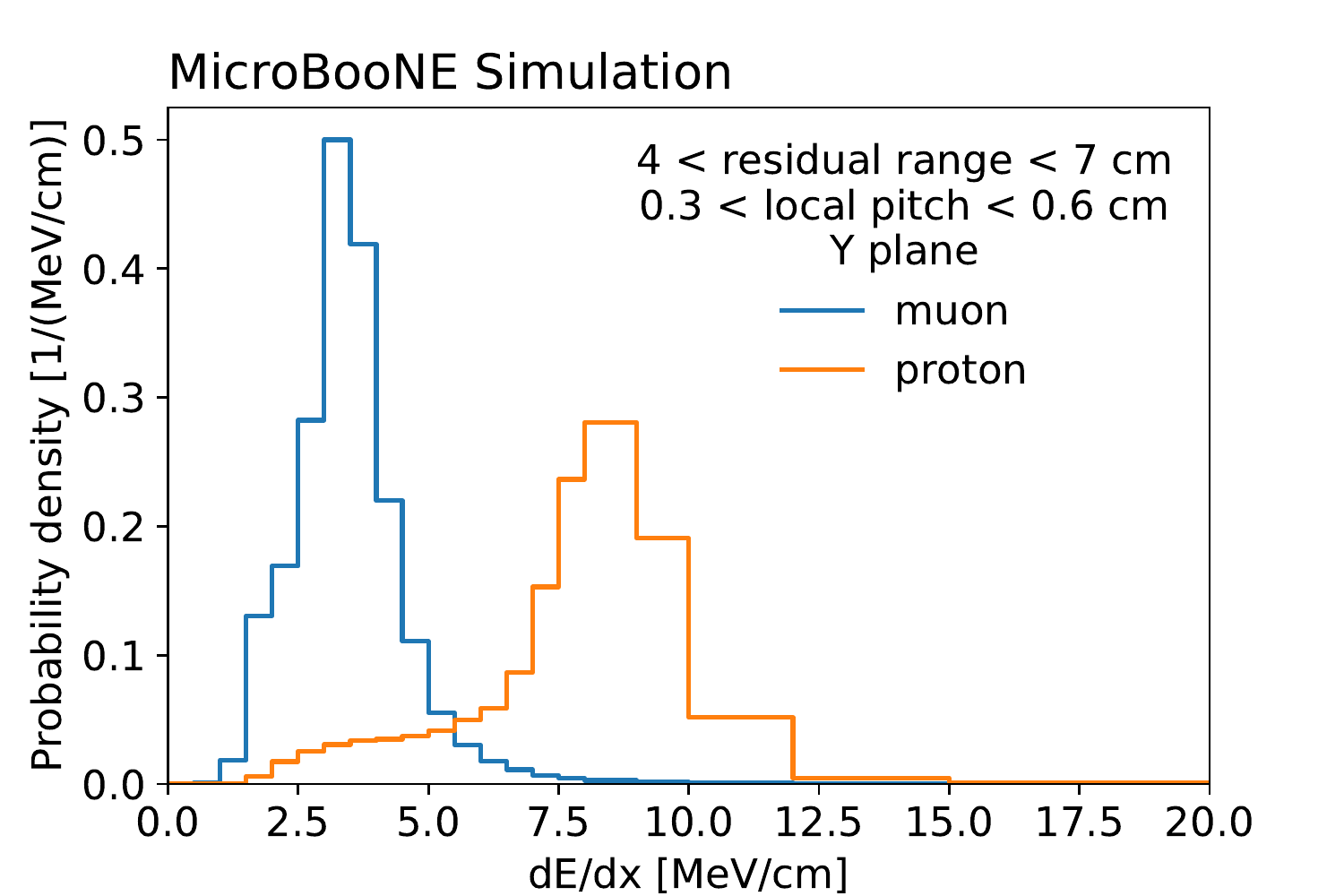}}
\subfloat {\includegraphics[width=0.33\textwidth]{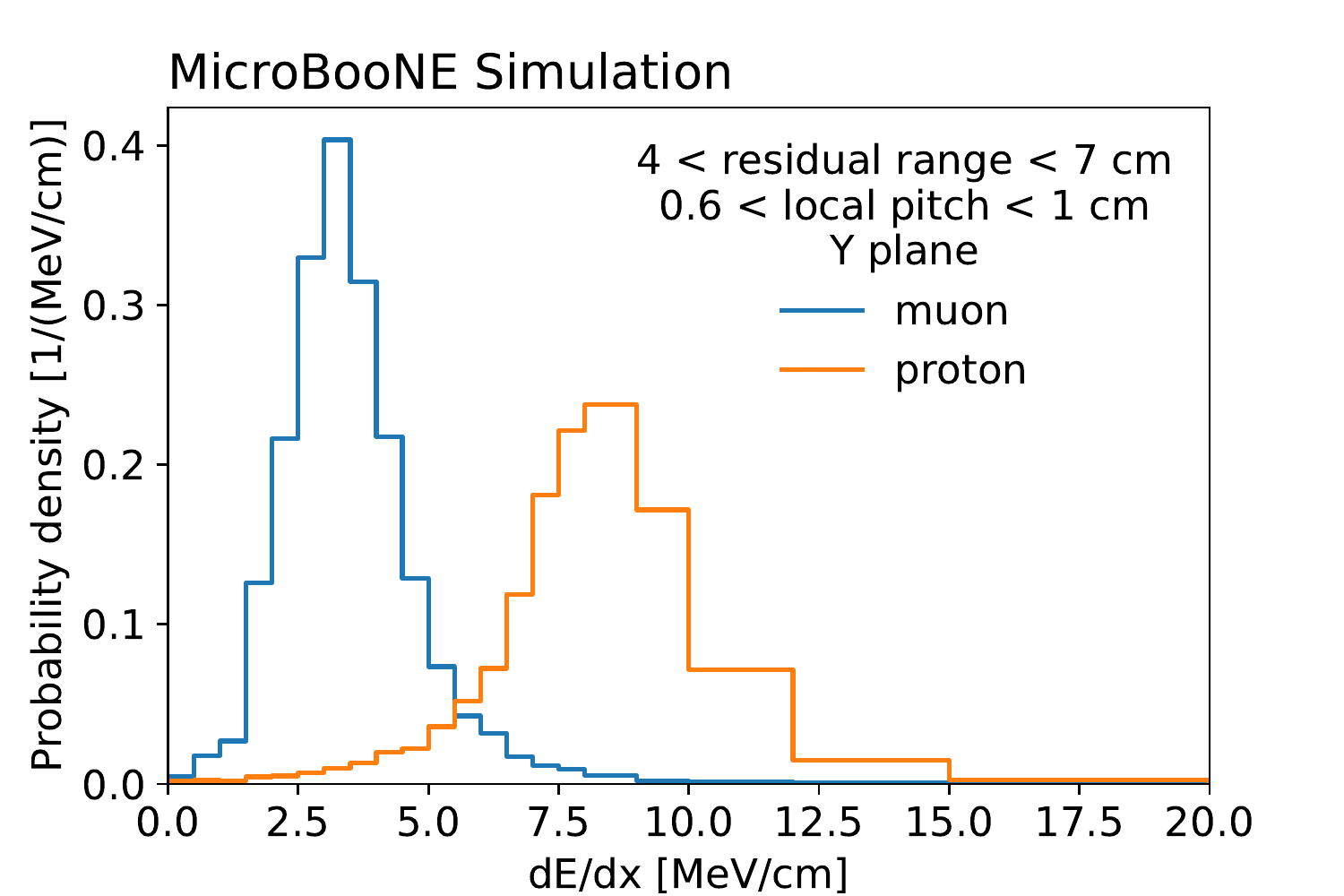}}
\subfloat {\includegraphics[width=0.33\textwidth]{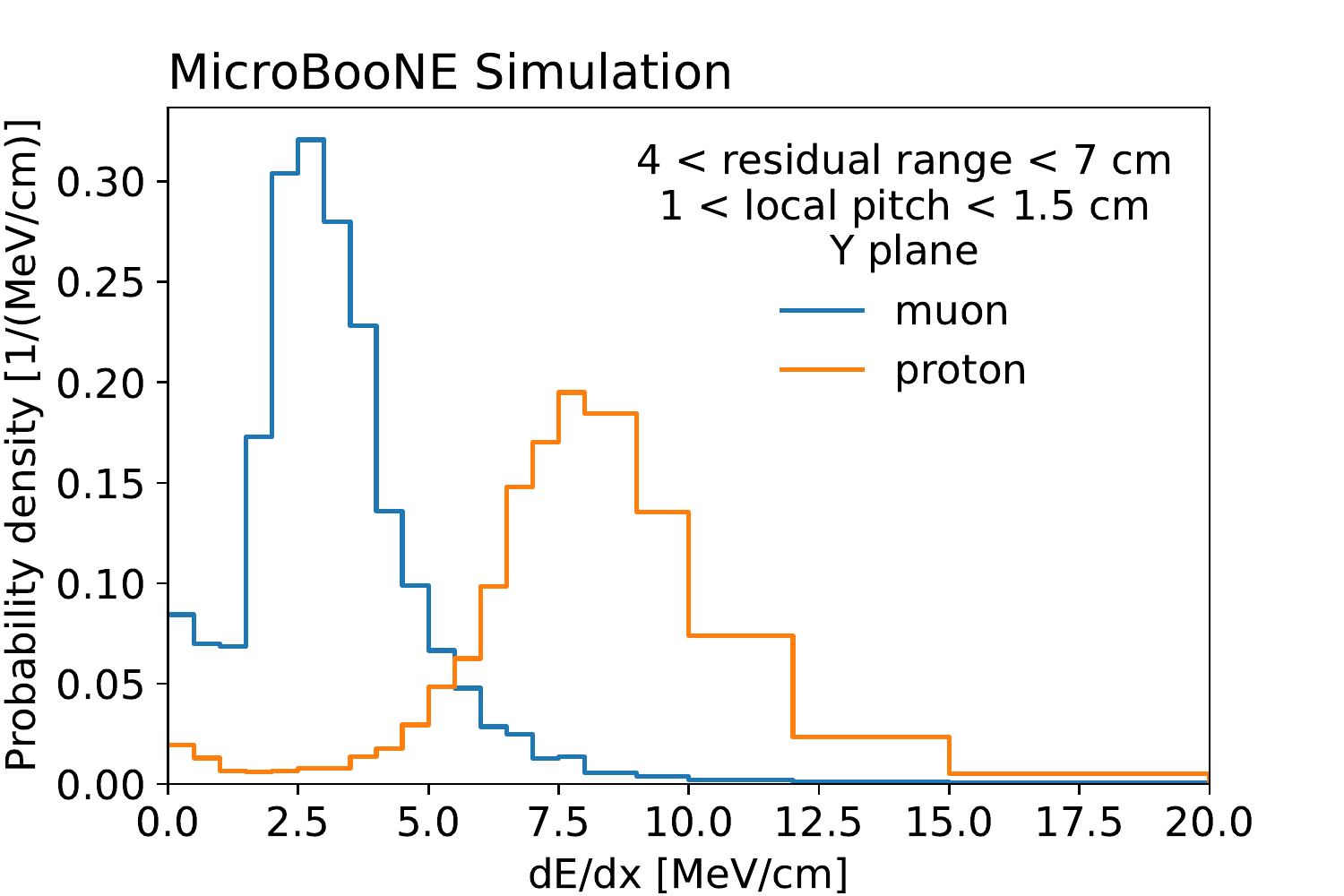}}

\subfloat {\includegraphics[width=0.33\textwidth]{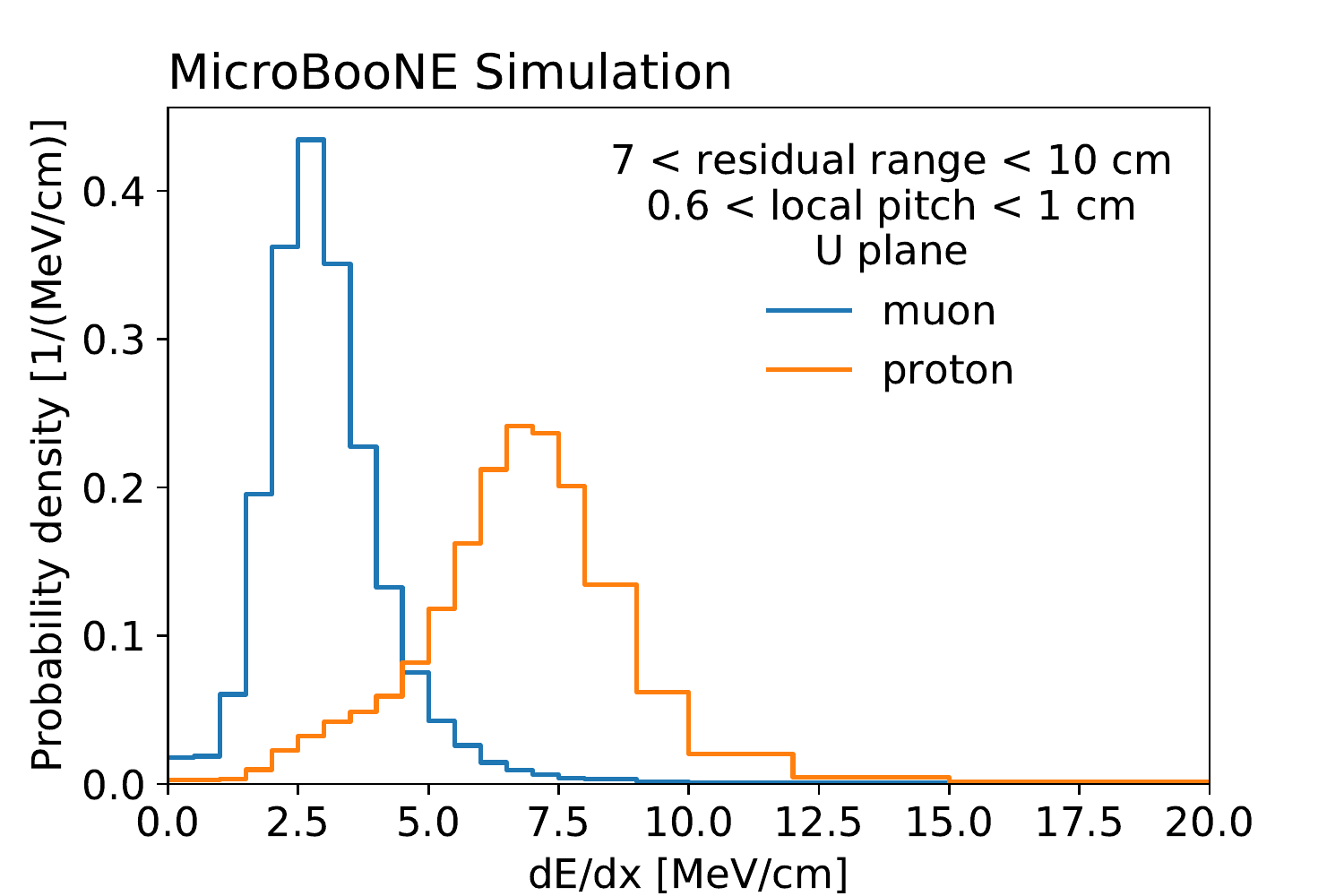}}
\subfloat {\includegraphics[width=0.33\textwidth]{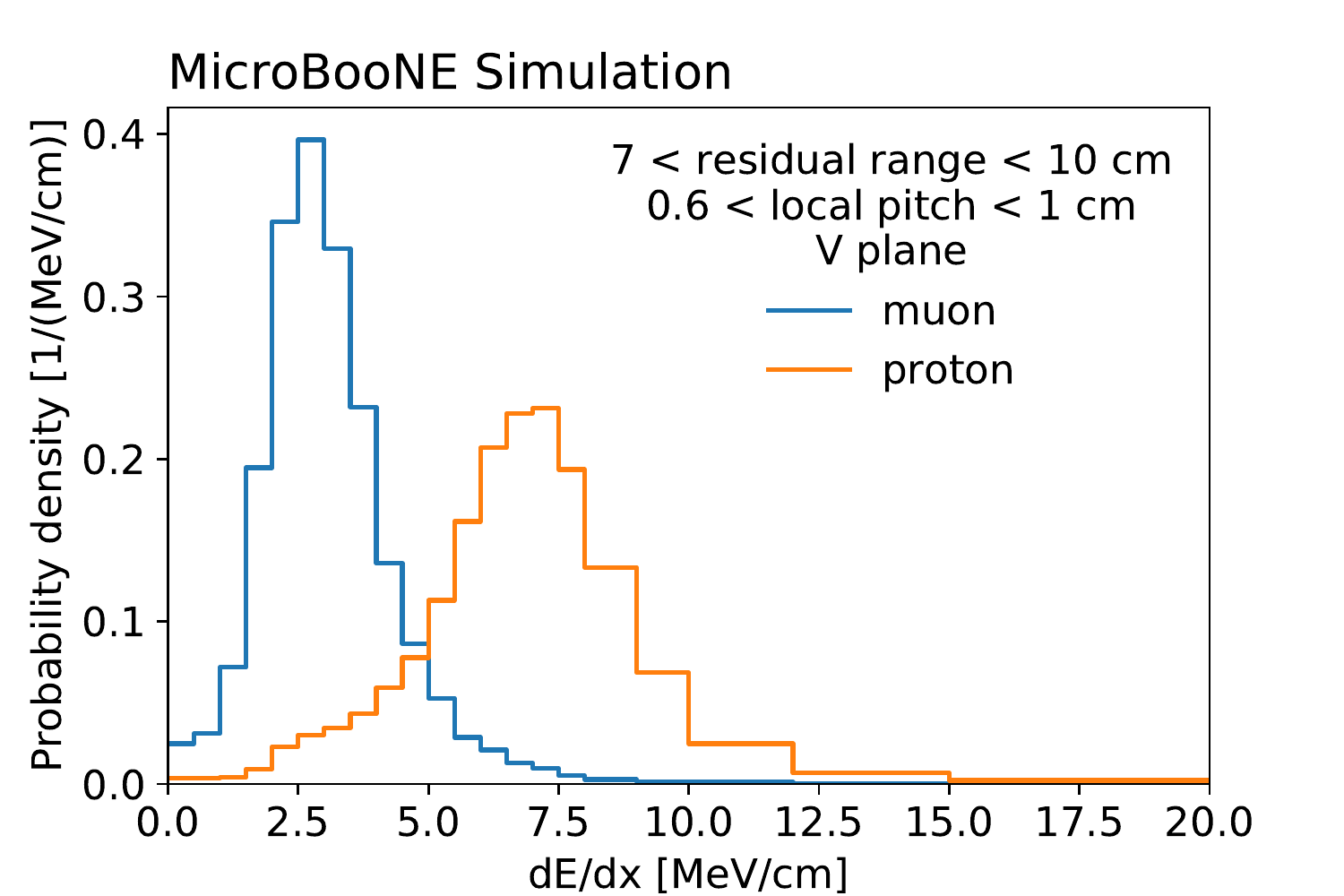}}
\subfloat {\includegraphics[width=0.33\textwidth]{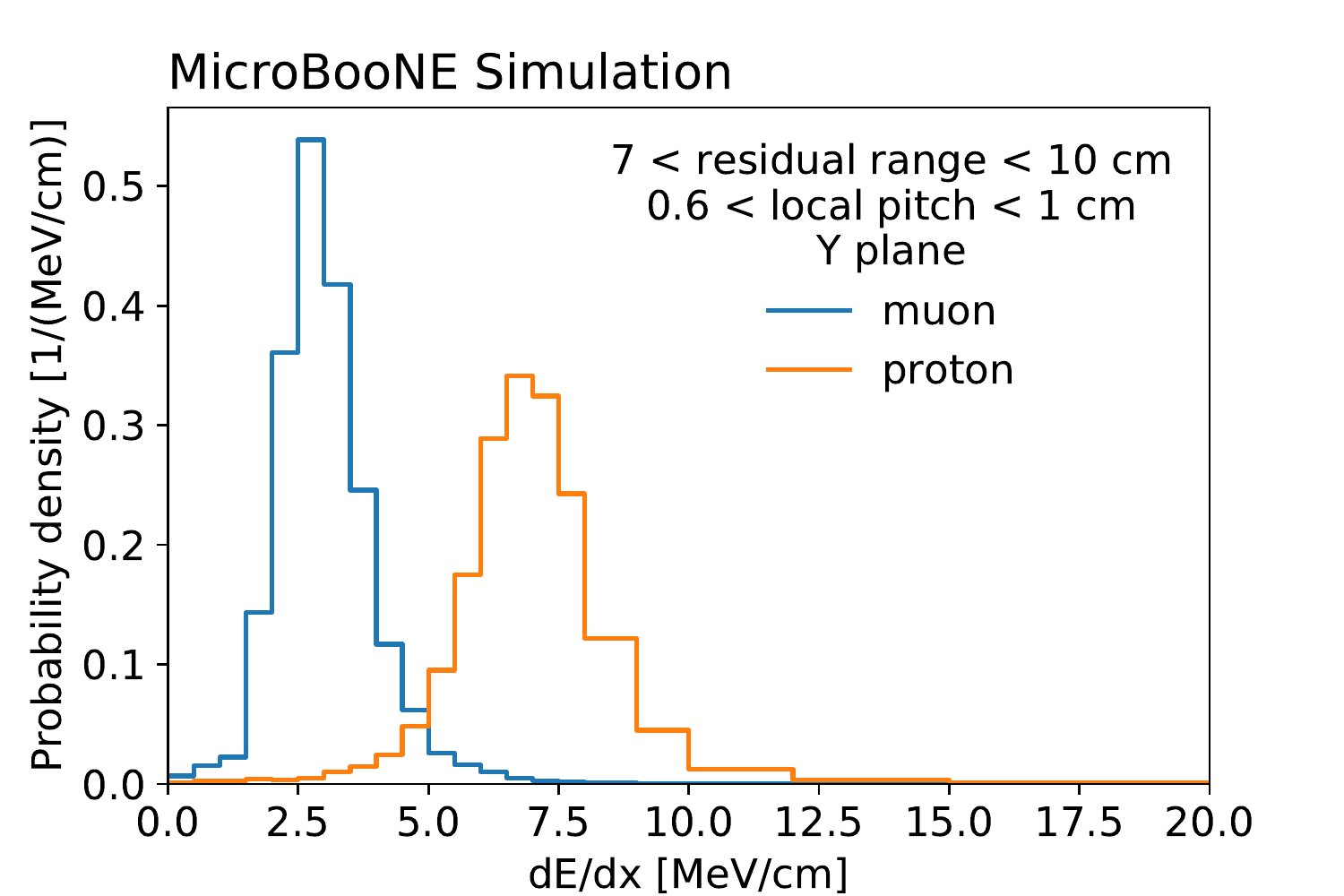}}
\caption{Expected \dedx distributions for muon (blue) and proton (orange) hits.
The top row shows distributions on the collection plane, with a fixed value of \pitch, and three different values of the residual range.
The middle row shows distributions on the collection plane, with a fixed value of the residual range, and three different values of \pitch.
The bottom row shows distributions on the three wire planes, with a fixed value of residual range and \pitch.
As expected, the peak \dedx value reduces at higher residual range, passing from 11 MeV/cm to 5.75 MeV/cm, and to 4.25 MeV/cm across the three bins under consideration for protons.} 
\label{fig:llr_pid_pdf_example}
\end{figure}

The PDF is estimated through a three dimensional histogram of \dedx, residual range, and \pitch.
The histogram is normalized so that for each combination of values of residual range and \pitch, the integral of the \dedx distribution sums to one, providing an estimate of the conditional PDF $p(\differential E / \differential x | \text{residual range}, \text{\pitch})$ that is not informed by the underlying kinematics of tracks.
This procedure is repeated for each plane and for two particle species, namely muons and protons.
The histograms are filled with hits associated with well-reconstructed tracks produced in simulated neutrino interactions.
These tracks are required to be complete, meaning that more than 90\% of the true deposited charge is reconstructed.
They are also required to be pure, meaning that more than 90\% of their reconstructed charge was deposited by a single particle.
The tracks are also required to be contained within a fiducial volume, where both the start and end points are at least 20 cm away from the boundaries of the TPC.

The \dedx PDF is visualized through three series of examples in \cref{fig:llr_pid_pdf_example}, where in each row only one of the three parameters (residual range, \pitch, and plane, respectively) is varied, while keeping the other two fixed.
The PDF changes considerably, showing, for example, a reduction of the \dedx of the peak value at higher residual range and an increase of the width at higher \pitch, justifying the need for such a construction as a function of these three variables.

\subsection{The likelihood ratio test statistic as PID score}
Using the PDF previously constructed, the likelihood of any particle hypothesis can be computed for each reconstructed track.
Interactions of neutrinos in the GeV energy range in liquid argon lead to comparable rates of muons, charged pions, and protons, making the classification between these particle species important.
However, this paper will focus on the binary classification problem of distinguishing muons from protons.
As pions and muons have very similar masses, the calorimetric separation of these two particle species is not addressed in this paper, and pion tracks will appear as muon-like by means of this algorithm.
Kaons are instead very rare (approximately 0.1\% of the events in \ub are predicted to contain a kaon) and are omitted in this work.

The likelihood for a track is computed starting from the single-hit-likelihood:
\begin{equation}
\label{eqn:likelihood_hit}
    \mathcal{L}_{\text{hit}}(\text{type} | \text{plane}, dE/dx, \text{rr}, \text{\pitch}) = p(dE/dx | \text{type}, \text{plane}, \text{rr}, \text{\pitch}),
\end{equation} 
where $p$ stands for the PDF, type refers to muon or proton and rr stands for residual range. 
The \pitch is measured locally, and it is generally different for each hit associated with the same track, because of changes in track trajectory due to multiple Coulomb scattering.
The plane is included because the PDFs are significantly different for the different planes.
The single-plane-likelihood is computed by taking the product of the single-hit-likelihood for each hit on a given plane:
\begin{multline}
\label{eqn:likelihood_plane}
        \mathcal{L}_{\text{plane}}(\text{type} | \text{plane}, \{dE/dx\}_{i = 1, ..., N}, \{\text{rr}\}_{i = 1, ..., N}, \{\text{\pitch}\}_{i = 1, ..., N}) = \\
        \prod_{i=1}^N \mathcal{L}_{\text{hit}}(\text{type} | \text{plane}, dE/dx_i, \text{rr}_i, \text{\pitch}_i),
\end{multline}
where $i = 1, ..., N$ indexes the hits on the plane under consideration.
The three-plane-likelihood, which is the likelihood for the entire track, is then computed as the product of the single-plane-likelihoods for the three wire planes.

The likelihood defined this way is an approximation as it neglects correlations between the charge measured on different wires and planes.
However, fluctuations of the hit charge are in general correlated among wires on different planes, as they record the same charge through different projections.
Moreover, the induced charge on neighboring wires and correlated noise introduce additional correlations between the charge recorded on different wires on the same plane \cite{ub_signal_processing1, ub_signal_processing2}.
Modeling these correlations is in general complex, as they depend on the geometry on a track-by-track basis.
Neglecting such correlations and using an approximation of the likelihood makes the method less optimal, and may result in a loss of separation power.
A possible discrepancy between the data and the simulation for the values of \dedx, which are the inputs of the PID method, could introduce a systematic bias.
However, as shown in the plots in \cref{sec:applications} where the data and the simulation are compared, there is no evidence that this effect is important.
In fact, all the correlation and noise effects introduced previously are reproduced in the simulation \cite{ub_signal_processing1, ub_signal_processing2}, and a dedicated correction of the \dedx distribution in angular bins is applied on top of the overall calibration \cite{ub_calibration}, making the simulation precise and accurate.


The likelihood is then used to compute the likelihood ratio test statistic, which is employed to perform the classification task:
\begin{gather}
\label{eqn:likelihood_ratio}
T(dE/dx, \text{rr}, \text{\pitch}) = \mathcal{L}(\text{muon}| dE/dx, \text{rr}, \text{\pitch}) /  \mathcal{L}(\text{proton}| dE/dx, \text{rr}, \text{\pitch}),
\end{gather}
where the indices running on wires and planes have been omitted here for simplicity, and either a single-plane-likelihood (\cref{eqn:likelihood_plane}) or the three-plane-likelihood can be considered.

The binary classification problem of distinguishing protons from muons has the likelihood ratio as the most powerful statistical test, as proven by the Neyman-Pearson lemma.
It provides the largest classification efficiency for any given value of the mis-identification rate.

For computational purposes, in the rest of article, instead of $T$, the PID score \pid will be considered, defined as:
\begin{equation}
    \mathcal{P} = \frac{2}{\pi} \arctan{\left(\log(T)/100\right)}.
\end{equation}
Computing the logarithm of $T$ is convenient as it reduces to a sum of log-likelihoods rather than a product of likelihoods.
This bijective non-linear transformation of $T$ does not change the separation power of the method, but it constrains the value of the PID score \pid, otherwise unbounded, between -1 and 1, making it easier to display.

\subsection{Performance of the particle identification}
\label{sec:performance}
The performance of \pid is evaluated on a test sample of more than 20000 simulated protons and muons, selected in the same manner as in \cref{sec:pdf}.
The sample contains inclusive neutrino interactions from the Booster Neutrino Beam (BNB) simulated with GENIE v3.0.6 \cite{genie}, using a tailored MicroBooNE tune \cite{genie_tune}.
\begin{figure}[!h]
    \centering
    \subfloat{\includegraphics[width=0.5\textwidth]{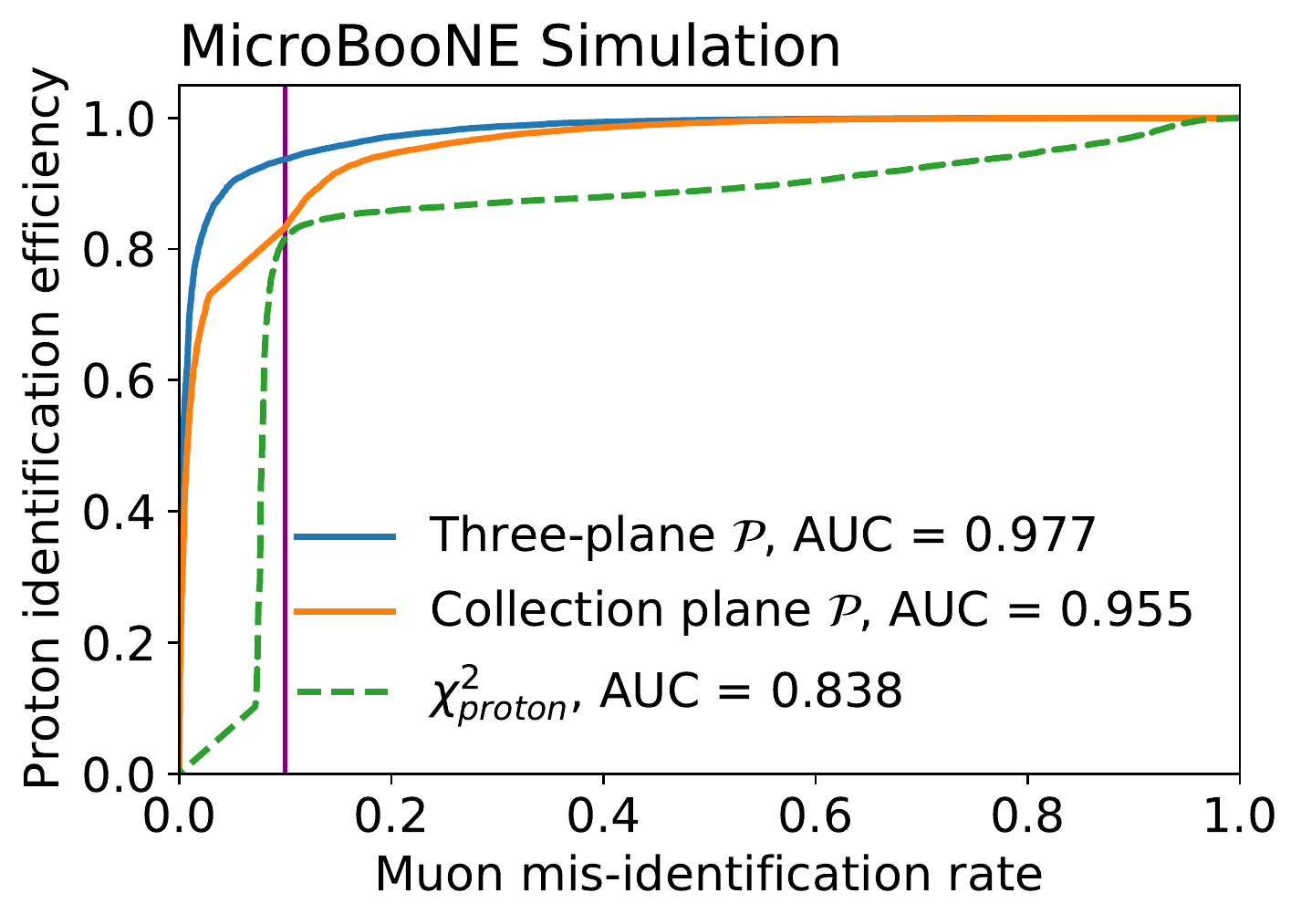}}
    \subfloat{\includegraphics[width=0.5\textwidth]{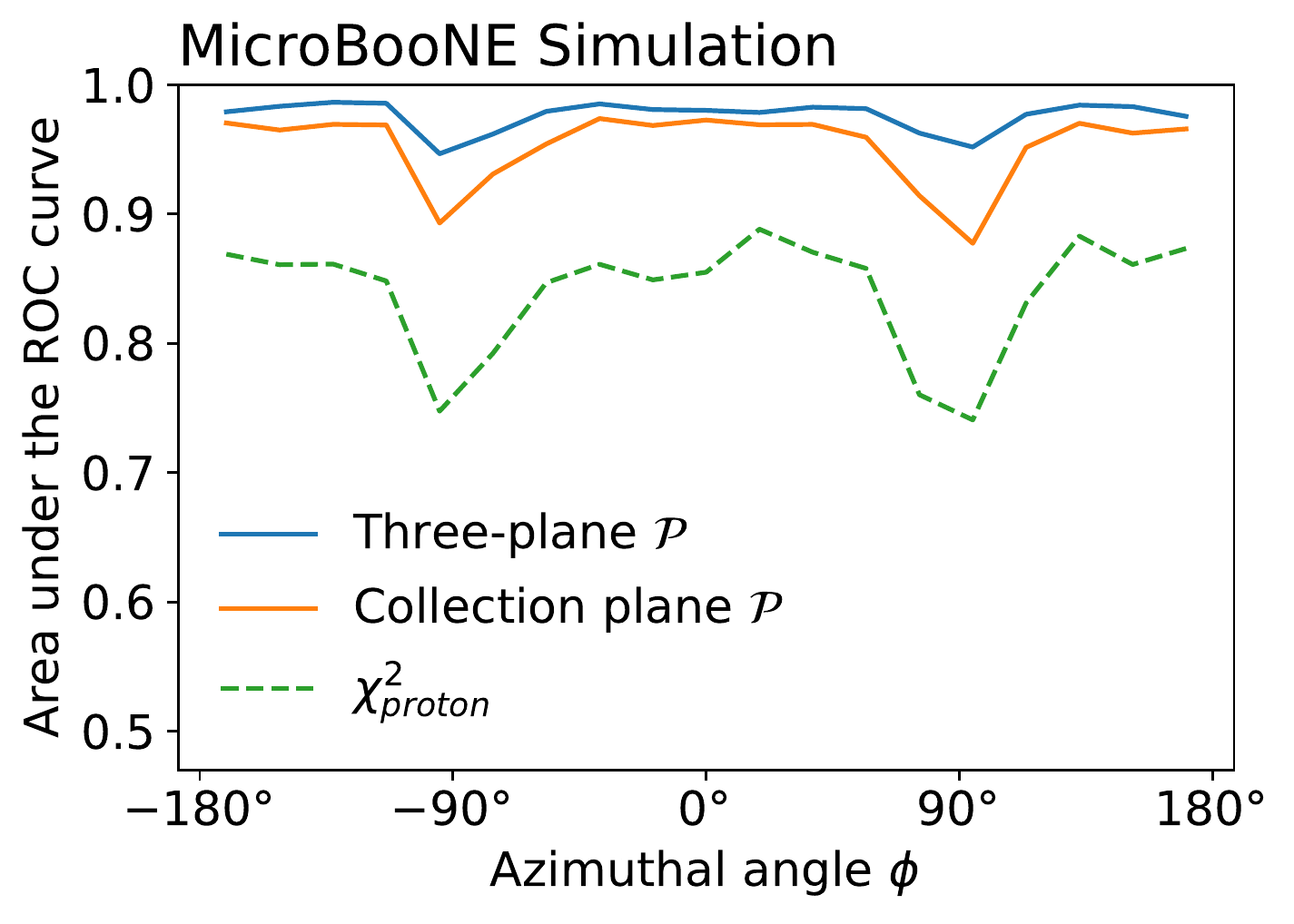}}
    \caption{Comparison of the proton/muon separation power of different PID scores.
    The left plot shows the ROC curves on the entire sample, and the right plot shows the area under the curve (AUC) in bins of the track angle $\phi$.
    The blue curves refer to the proposed PID score \pid using three planes, the orange refer to the collection plane only \pid, whereas the green curves show the $\chi^2$ test with respect to the proton hypothesis.
    The purple vertical line on the left plot slices the curves at muon mis-identification rate 0.1, comparing the proton identification efficiencies of the three methods at the same working point.
    } 
    \label{fig:roc_curves}
\end{figure}
A receiver operating characteristic (ROC) curve is calculated from the test statistics distributions for the two particle types and shown in the left plot of \cref{fig:roc_curves} for the three-plane \pid, for the collection plane only \pid, and for the $\chi^2$ test with respect to the proton hypothesis, which represents the previous state of the art \cite{ub_calibration}.
The latter quantity, computed by comparing the data with the expectation from Bethe-Bloch theory, has been used in several previous MicroBooNE analyses and it is shown here as a reference for comparison.
The ROC curves show the proton efficiency as a function of the muon mis-identification rate, which are bounded between 0 and 1.
For a given method, every possible cut value between -1 and 1 corresponds to a point on the ROC curve.
The performance is quantified at the working point with 10\% of the muons mis-identified as protons: the three-plane \pid provides 93.7\% efficiency at selecting protons compared to 83.4\% for the collection plane only \pid and 81.6\% for the $\chi^2$ test with respect to the proton hypothesis.
An overall measure of the separation power is defined using the area under the ROC curve (AUC).
When this metric is equal to 1, the variable allows perfect separation at any working point, whereas a value of 0.5 represents a random guess.
The three-plane \pid scores a AUC of 0.977 compared to 0.955 for the collection-plane only \pid and 0.838 for the $\chi^2$ test with respect to the proton hypothesis.
The robustness of the quoted performance is tested against detector systematic uncertainties.
The performance is evaluated on a series of simulations with a modified detector response to assess the detector systematic uncertainty.
This leads to an uncertainty on the proton efficiency of 1.2\% at 10\% muon mis-id.
The uncertainty on the AUC is 0.002 units for the nominal value of 0.976.
This uncertainty is dominated by the modeling of electron-ion recombination.
The statistical uncertainty on the efficiency and AUC determination is negligible.
The right plot of \cref{fig:roc_curves} shows the AUC in bins of the azimuthal angle $\phi$ of the track, which describes the direction of the track on the plane orthogonal to the beam direction: $\phi = 0^\circ, \pm 180^\circ$ refer to the drift direction and $\phi = \pm 90 ^\circ$ refer to the vertical direction.
Both plots illustrate an overall improvement of the separation power with respect to the $\chi^2$ test with respect to the proton hypothesis, and a mitigation of the dependence of the performance on the track angle.
Combining the three planes improves the separation power in every angular region, especially for vertical tracks ($\phi\sim\pm 90^\circ$), where the collection plane is the least effective.

\section{Applications to physics cases}
\label{sec:applications}

The following analyses were developed using data collected by \ub with the BNB during winter and spring 2016.
This data amounts to $4.8 \times 10^{19}$ protons on target (POT).
This data, in which neutrino interactions are present, is labeled as DATA Beam ON.
The prediction comes from a combination of the simulation of neutrino interactions and data collected out of the beam windows, labeled as DATA Beam OFF.
Even in events where a neutrino interaction is present, $\mathcal{O}(10)$ cosmic rays cross the detector on average.
Instead of being simulated, cosmic ray waveforms are acquired out of the beam window and overlaid to simulated neutrino interactions.

\subsection{Proton-muon separation for tracks recorded on data}
\label{sec:proton_muon_separation}
The first test performed is to verify if the result obtained in the simulation in \cref{sec:performance} holds also with neutrino data.
\begin{figure}[!ht]
    \centering
    \includegraphics[width=0.7\textwidth]{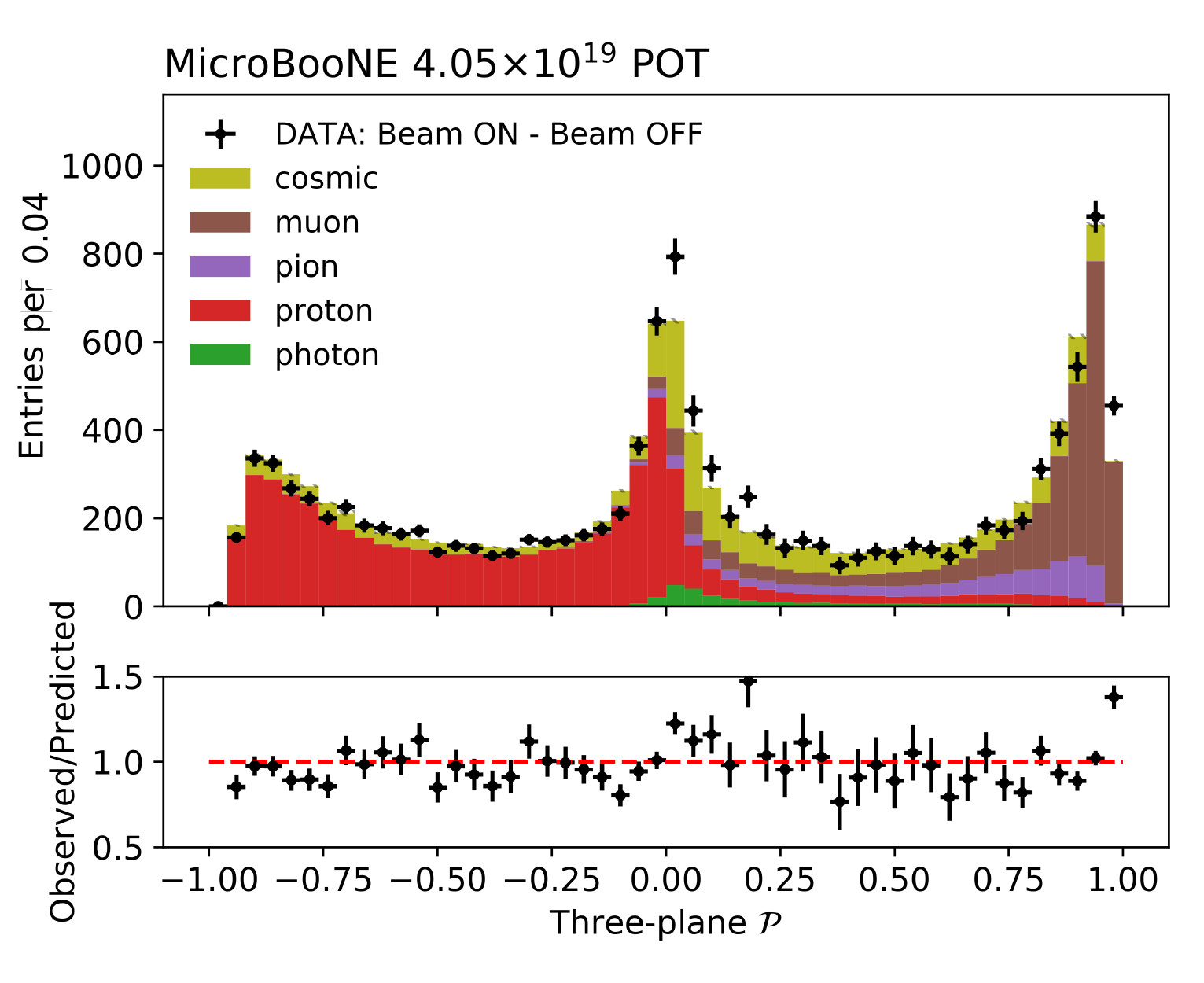}
    \caption{Distribution of the three-plane \pid for neutrino-induced tracks selected in data and simulation.
    The data (black cross) shows the difference between the DATA Beam ON and the DATA Beam OFF, in order to remove the contribution from non-beam events.
    The simulation (stacked histogram) is normalized to the same number of events observed in the data, and it is broken down for different particle species.
    In the case the particle selected in the simulation is an overlaid cosmic, it is assigned to the category "cosmic".
    The uncertainties shown on the data points are the expected statistical uncertainties from Poisson counting.
    }
    \label{fig:llr_pid_uvy_example}
\end{figure}
Tracks are selected by requiring track-score\,$>$\,0.5, a measurement of the likeliness of a reconstructed particle to be a track, with values ranging from 0 for shower-like particles to 1 for track-like particles.
Track-score is provided by the Pandora reconstruction.
Tracks are also required to be reconstructed within 5\,cm from the vertex, and to be contained within a fiducial volume, defined as the set of points that are at least 20 cm apart from every side of the TPC.
\Cref{fig:llr_pid_uvy_example} shows the distribution of the PID score for these tracks, comparing the data (black cross) with the simulation (stacked colored histogram).
Protons, reconstructed with a low \pid, populate the left side of the distribution. 
These are well separated from lighter particles, such as muons and pions, which populate the region at larger values of \pid. 
Tracks associated with cosmic rays are distributed along the whole spectrum, as they can be induced by cosmic muons or by protons kicked out of the argon nuclei.
A peak at \pid $\sim 0$ is also present.
These are short tracks, for which there is too little information to discriminate between the two hypotheses. 
In fact, $\log(T)$ is additive for each hit: the longer the track, the more hits, the larger $\log(T)$ and thus \pid can be.
The simulation reproduces the shape of the data, confirming the performance studied in the simulation.
\begin{figure}[!ht]
\centering
\subfloat{\includegraphics[width=0.5\textwidth]{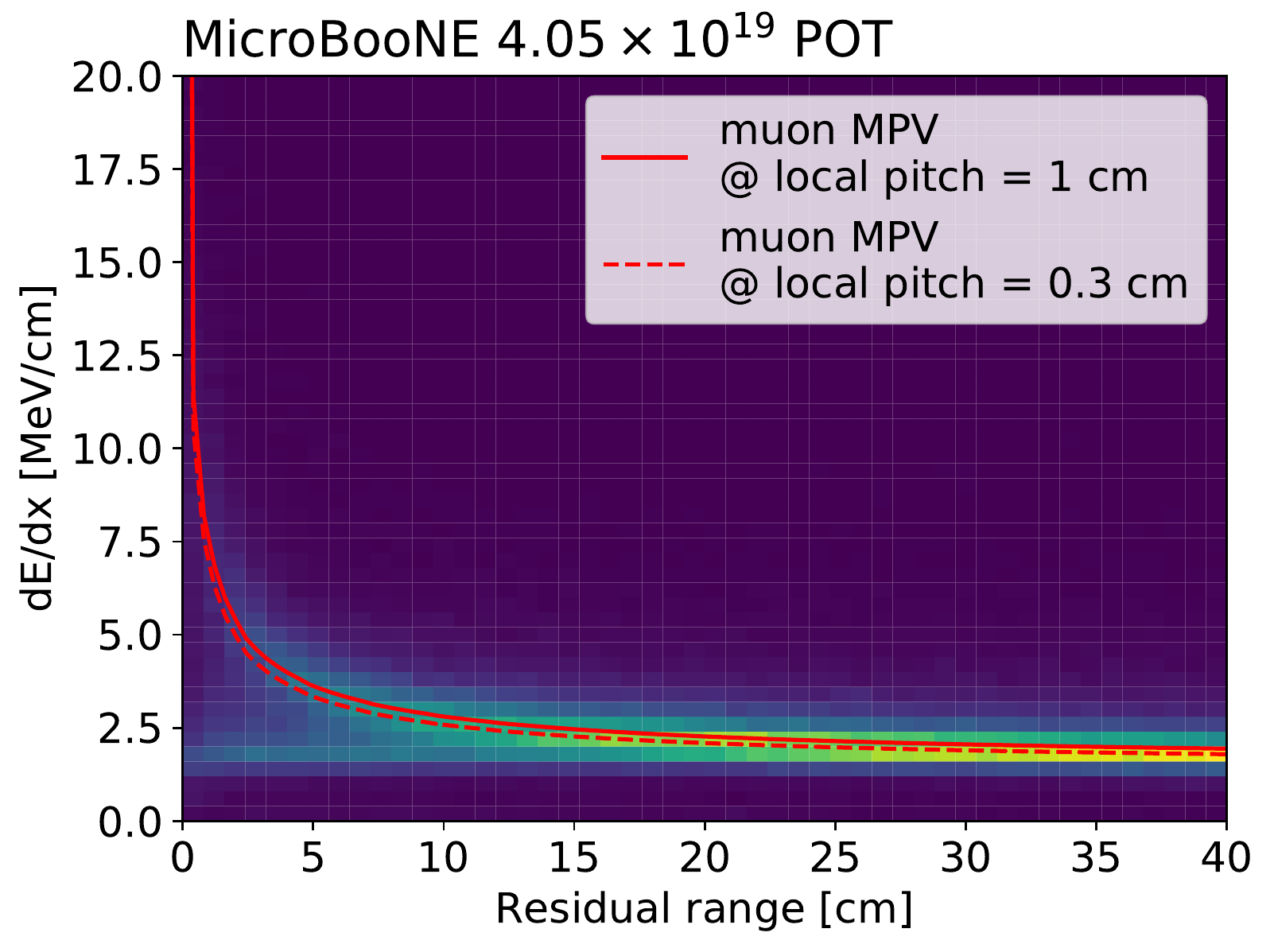}}\hfill
\subfloat{\includegraphics[width=0.5\textwidth]{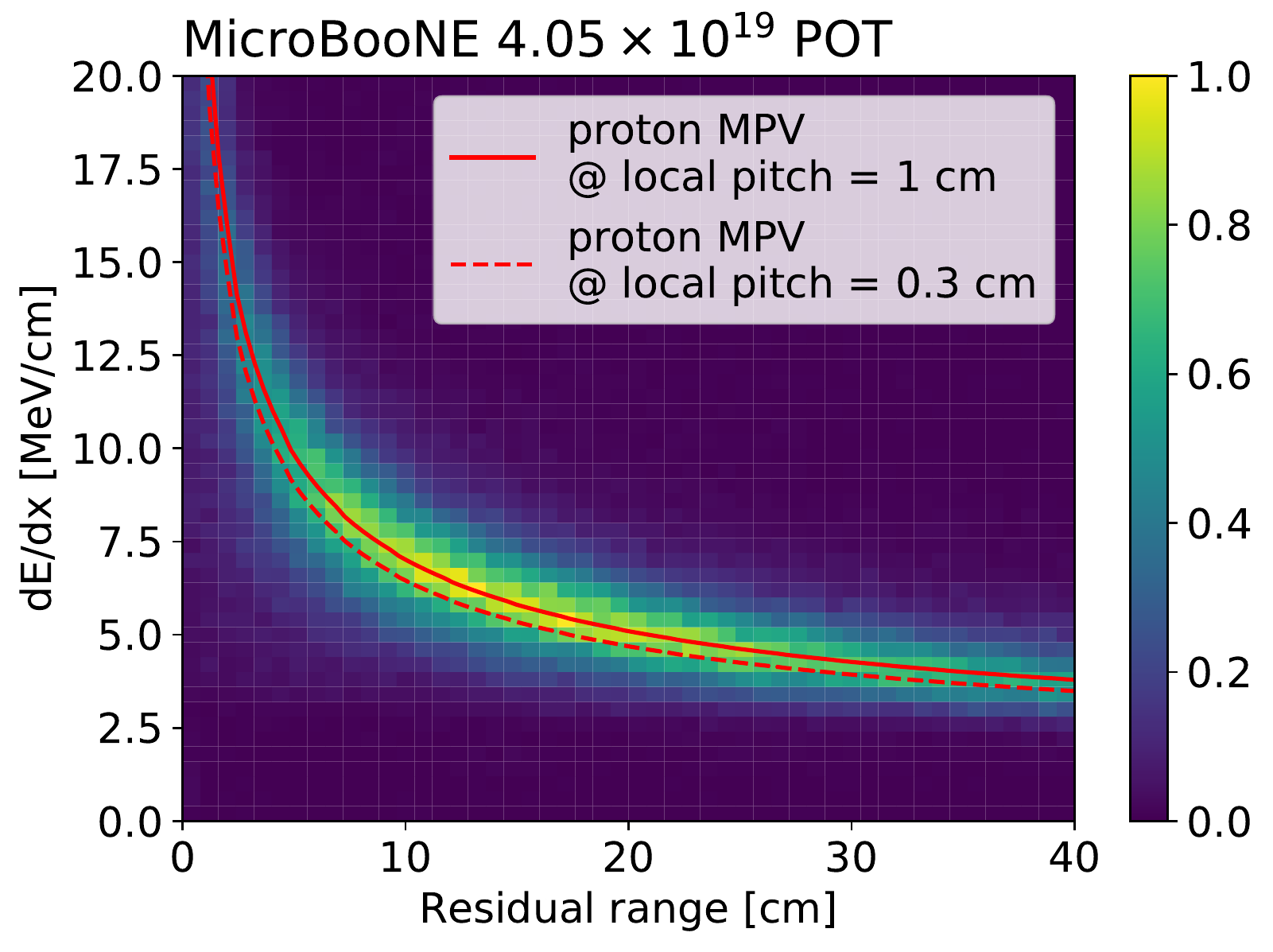}}
\caption{Collection-plane \dedx vs residual range profile for well reconstructed, contained, and low \pitch tracks in data, identified as muon (proton) candidates on the left (right) plot. 
The profiles are compared to the most probable values (MPV) as expected by the theory for the extremes of the range of \pitch under consideration (red lines).
The two plots are normalized to the maximum value in order to share a common color scale.}
\label{fig:dedx_vs_rr}
\end{figure}
Two additional plots (\cref{fig:dedx_vs_rr}) illustrate that \pid correctly identifies the Bragg peaks, in good agreement with the theoretical prediction.
Fully contained tracks, with track-score\,$>$\,0.8 and collection plane \pitch\,$<$\,1 cm are selected in beam data events.
The 2D distributions \dedx vs residual range on the collection plane, for muon-like tracks with \pid\,$>$\,0.5, and proton-like tracks with \pid\,$<$\,-0.5, are plotted on the left and right of \cref{fig:dedx_vs_rr}, respectively.
The two Bragg peaks are clearly visible and distinct.
This is possible because of the track \pitch requirement: selecting hits with small \pitch ensures \dedx is measured properly, resulting in physical and meaningful values. 
The solid and dashed red lines show the theoretical prediction of the most probable value of the \dedx distribution for the extremes of the range of \pitch under consideration.
The core of the data distribution lies between the two bands, demonstrating good calorimetric reconstruction for small \pitch.

\subsection{Large collection-plane-\pitch tracks identified with the two induction planes}
The following example illustrates the efficacy of combining the calorimetric measurements performed with the three wire planes.
\begin{figure}[!ht]
\centering
\subfloat[First induction plane (U).] {\includegraphics[width=0.5\textwidth]{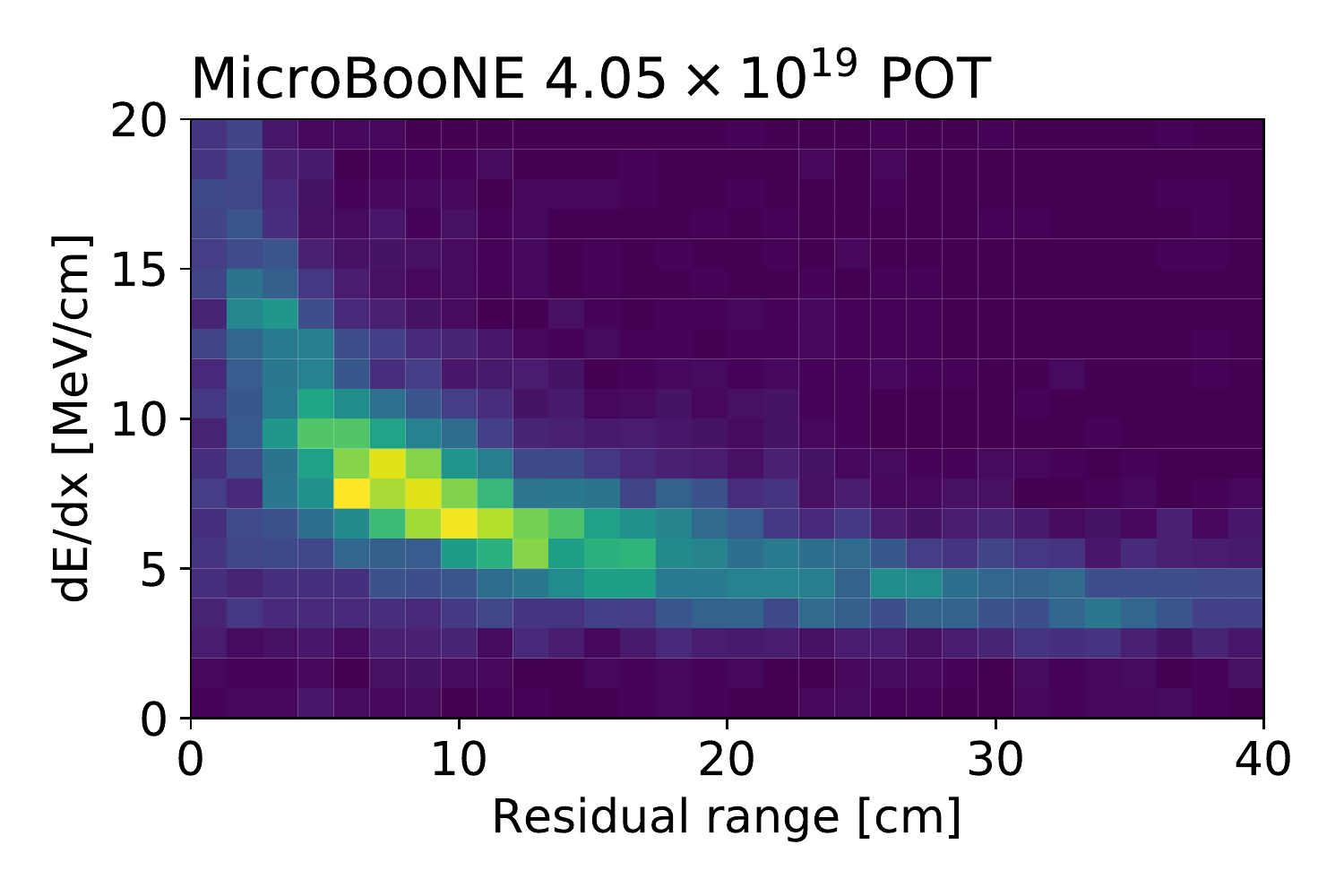}}\hfill
\subfloat[Second induction plane (V).] {\includegraphics[width=0.5\textwidth]{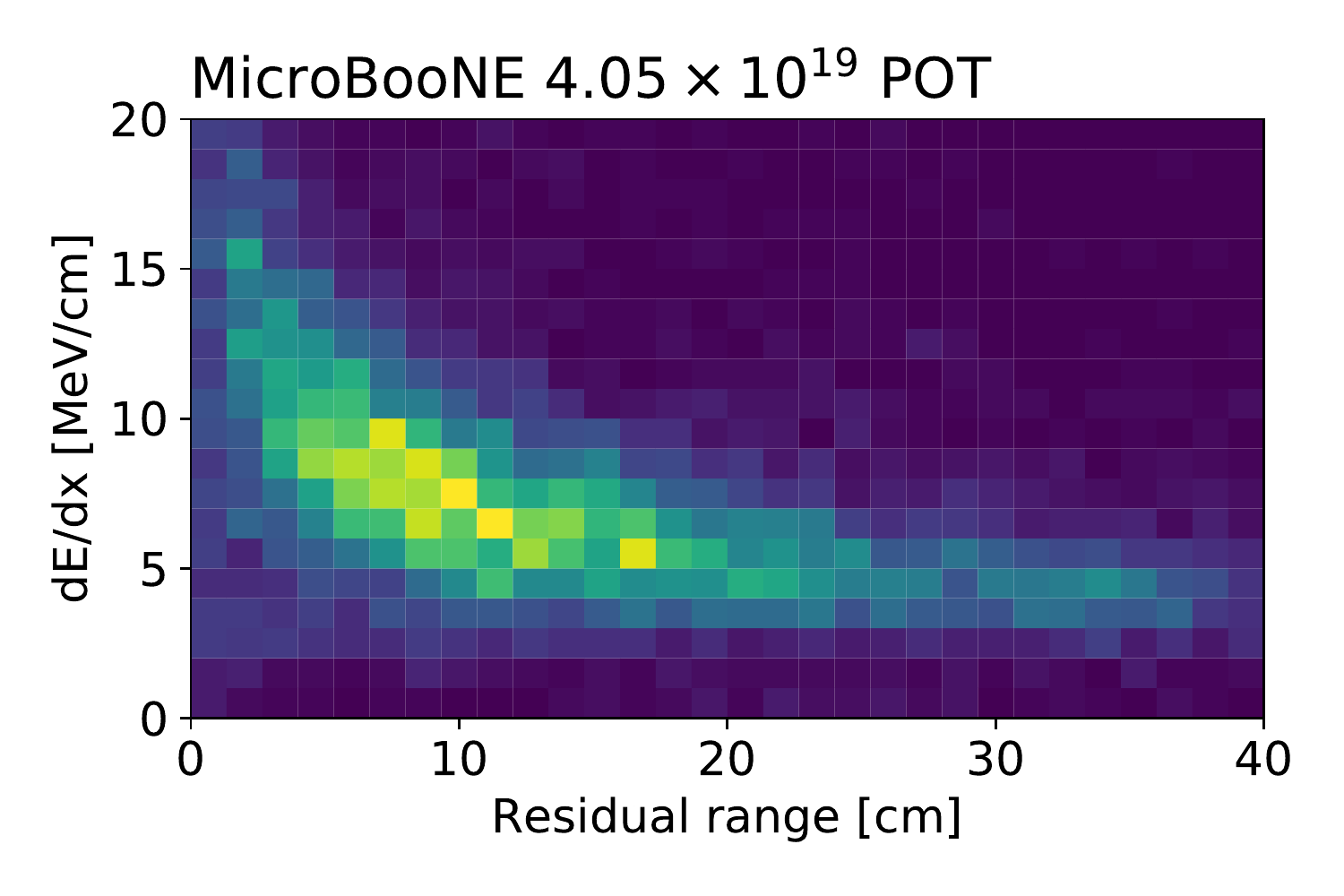}}\hfill
\subfloat[Collection plane (Y).] {\includegraphics[width=0.5\textwidth]{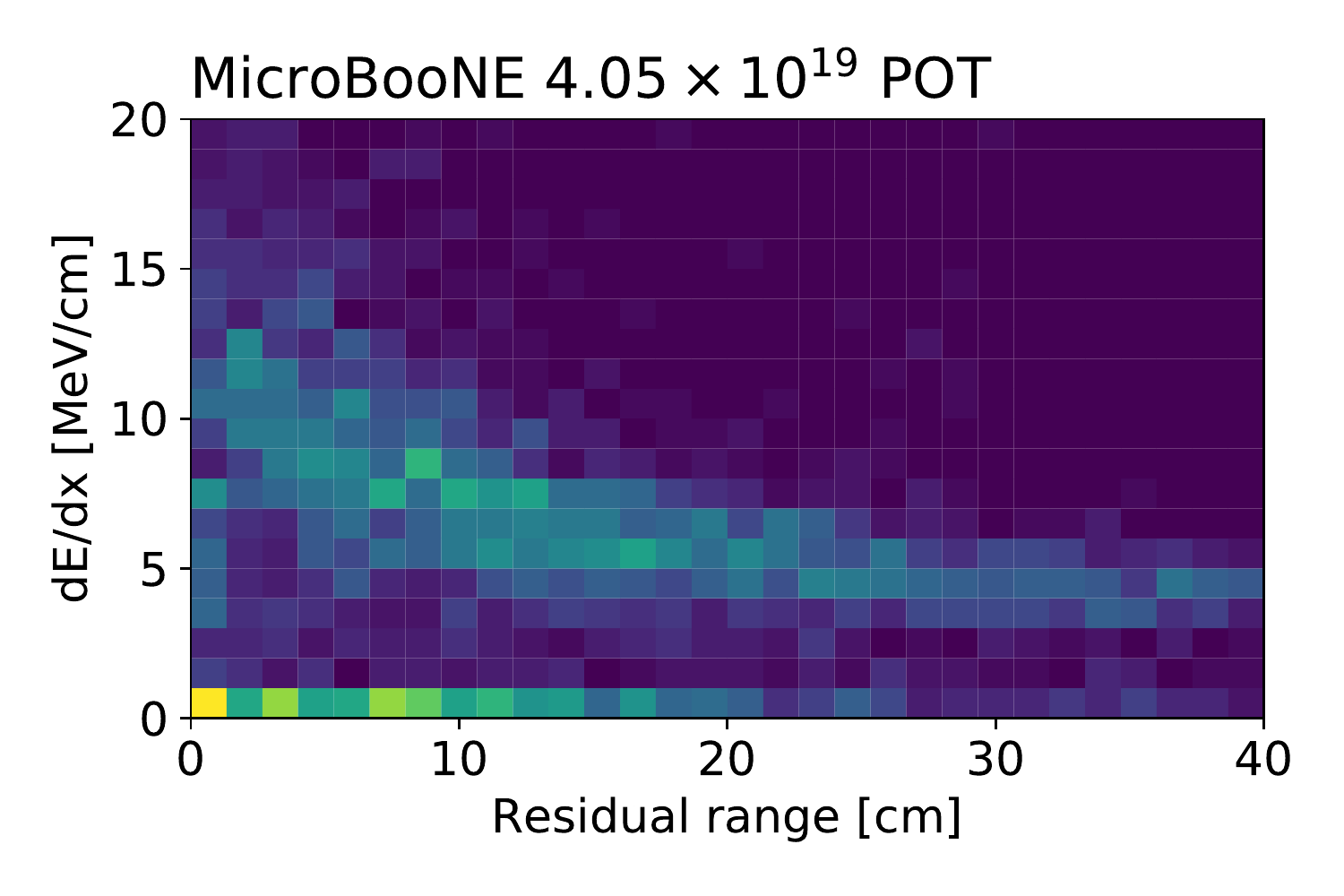}}
\caption{2D distribution of \dedx and residual range measured on the three wire planes for tracks identified as proton candidates in the data, with large collection plane \pitch.
}
\label{fig:dedx_vs_rr_high_pitch}
\end{figure}
\Cref{fig:dedx_vs_rr_high_pitch} shows the 2D distribution of \dedx and residual range measured on the U, V, and Y planes, for proton candidate tracks with large collection plane \pitch.
Proton candidates are required to be fully contained, and to have track-score\,$>$\,0.8. 
Proton-likeness is required through \pid$<$\,-0.5.
The collection plane \pitch is required to be larger than\,1 cm: such tracks lie on the plane orthogonal to the beam, traveling in directions where the calorimetric reconstruction is subject to large distortion.
For this set of tracks, only the induction planes exhibit the expected Bragg peak: combining the three planes recovers the separation power by correctly classifying protons whose calorimetric reconstruction is not accurate on one or more views.

\subsection{Exclusive \texorpdfstring{\numu}{numu} selection}

To further illustrate the separation power of \pid, a general \numucc selection targeting contained events is performed, and the selected events are classified into different exclusive channels.
Events are selected similarly to the procedure in \cite{numu_cross_section_microboone}, adding a containment requirement for all tracks reconstructed in the event, by requiring the start and end points of each track to lie inside the fiducial volume, as described in \cref{sec:proton_muon_separation}.
First, a muon candidate is chosen among the tracks attached to the vertex that are longer than 10 cm, by selecting the one with the largest \pid value.
\begin{figure}[!ht]
\centering
\subfloat {\includegraphics[width=0.55\textwidth]{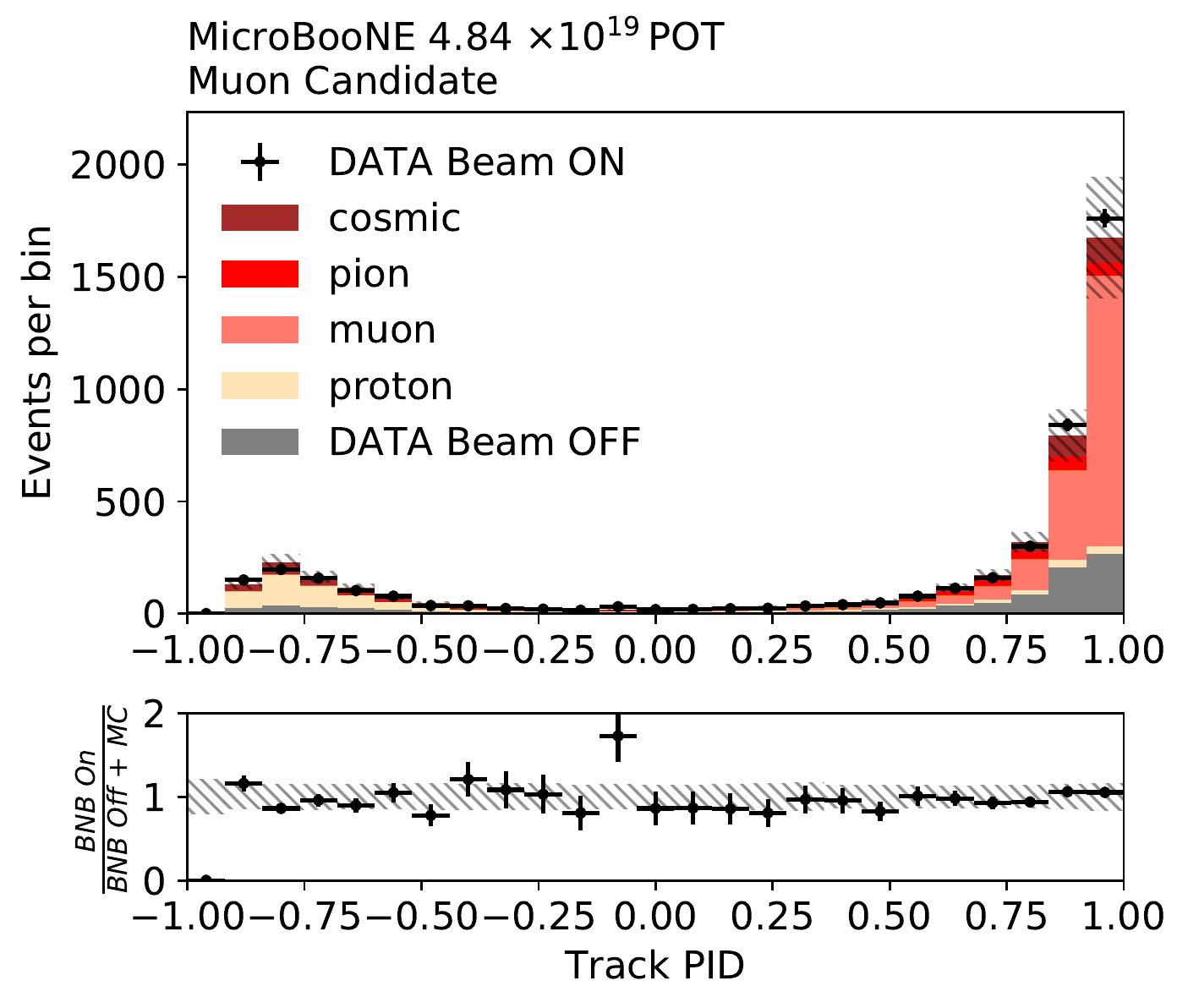}}\hfill
\subfloat {\includegraphics[width=0.2\textwidth]{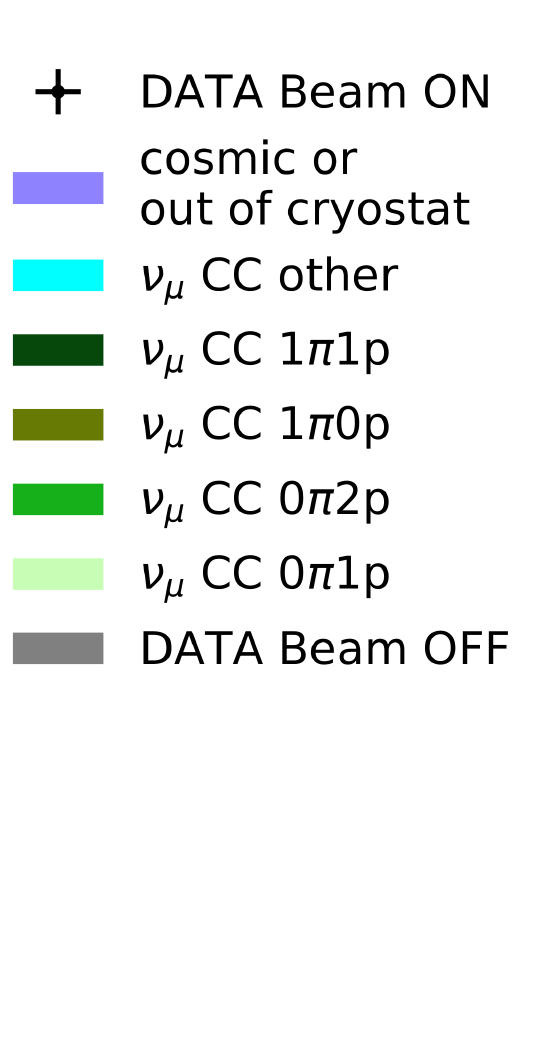}}\hfill
\subfloat {\includegraphics[width=\textwidth]{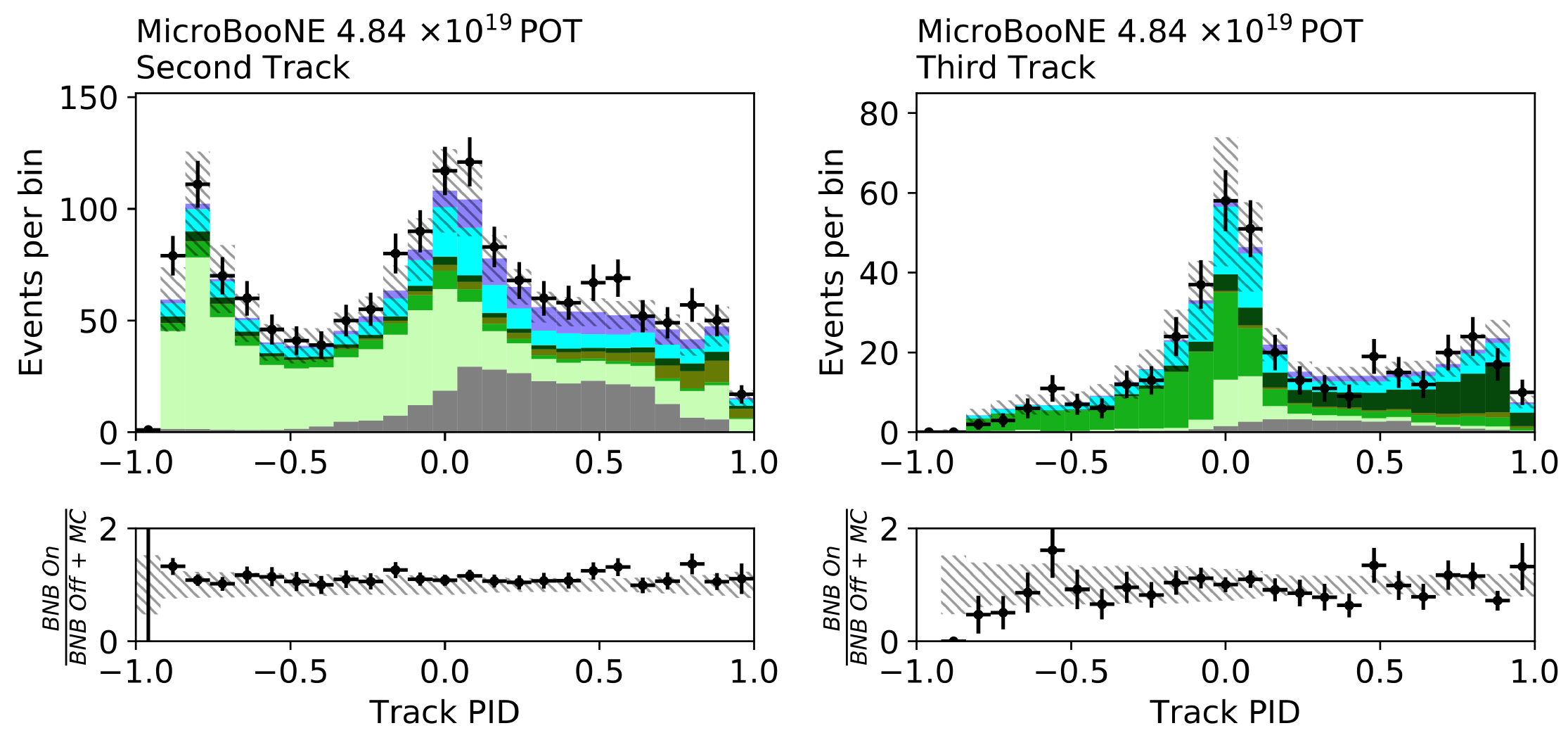}}\hfill
\caption{PID score distributions for the muon candidate track (top), the second track in events with two reconstructed tracks (bottom left), and the third track in events with three reconstructed tracks (bottom right).
The DATA Beam ON (black cross) is compared with the prediction based on the sum of the simulation of neutrino interactions (stacked histogram) and DATA Beam OFF (gray bars).
The selections are based on reconstructed quantities, while the prediction is broken down into different categories based on truth information.
In the first plot the different colors correspond to different particle types, while in the other two they correspond to different final states.
The uncertainties shown on the data points are the expected statistical uncertainties from Poisson counting, while the hashed patches on the stacked histogram illustrate systematic uncertainties on the prediction related to the simulation of the neutrino flux and interaction model. 
}
\label{fig:muon_selections}
\end{figure}
The top plot of \cref{fig:muon_selections} shows the distribution of \pid for muon candidates, showing a good separation between muon and proton tracks.
Selecting only events with a candidate with \pid $>$ 0.2 rejects most of the proton background, ensuring a pure selection of \numucc interactions.

Among the \numucc candidates, events with one additional reconstructed track (two-track events) are selected.
If correctly reconstructed, they result predominantly from \numucc interactions with either one proton and no pions ($\nu_{\mu} \text{CC} 0\pi 1\text{p}$) or one pion and no protons ($\nu_{\mu} \text{CC} 1\pi 0\text{p}$) in the final state.
In general, the former case predominantly (but not solely) results from quasi-elastic interactions while the latter is largely produced by the decay of a $\Delta$\, resonance.
The PID score of the second track (bottom left plot in \cref{fig:muon_selections}) separates the two cases, with $\nu_{\mu} \text{CC} 0\pi 1\text{p}$ populating the left side while the $\nu_{\mu} \text{CC} 1\pi 0\text{p}$ are located at positive values, because pions are indistinguishable from muons with this variable.
By considering the events with \pid $\leq 0$, we obtain a sample of contained $\nu_{\mu} \text{CC} 0\pi 1\text{p}$ interactions with 61\% purity and 40\% efficiency.
By applying the reverse cut, \pid $> 0$, we have a background rejection of 98\%, which provides the basis for a selection of contained $\nu_{\mu} \text{CC} 1\pi 0\text{p}$ interactions.
For this signature, the large cosmic ray background requires additional tailored background rejection.
With a similar methodology, events with two additional reconstructed tracks (three-track events), are selected.
Events with two protons and no pions in the final state ($\nu_{\mu} \text{CC} 0\pi 2\text{p}$), predicted to be mainly induced by meson-exchange current interactions and final state effects, can be distinguished from events with one proton and one pion in the final state ($\nu_{\mu} \text{CC} 1\pi 1\text{p}$), produced by a resonance decay.
Because the presence of a proton, identified by a large negative \pid, is common to the two cases, the track with the largest PID score among the two additional tracks (bottom right plot in \cref{fig:muon_selections}) is used to discriminate between $\nu_{\mu} \text{CC} 0\pi 2\text{p}$, on the left, and $\nu_{\mu} \text{CC} 1\pi 1\text{p}$, on the right.
The cut \pid $\leq 0$ provides a sample of contained $\nu_{\mu} \text{CC} 0\pi 2\text{p}$ interactions with 54\% purity and 24\% efficiency, while the reverse cut, \pid $> 0$, selects a sample of contained $\nu_{\mu} \text{CC} 1\pi 1\text{p}$ interactions with 25\% purity and 34\% efficiency.
In both cases the background rejection is over 99.5\%, emphasizing the difficulty of these selections, which could further benefit from additional cut variables.

Demonstrating the classification of exclusive \numucc final states is a novel result for liquid argon, and stems from the potential of the new PID score.
Future analyses will build on these examples, eventually leading to precise and detailed neutrino cross-section measurements.
\section{Conclusions}
\label{sec:conclusions}
The capability to perform precise calorimetric measurements is one of the most important factors that make LArTPCs powerful tools in the study of neutrino interactions.
This work illustrates a detailed study of the performance of the calorimetric reconstruction using MicroBooNE data, the longest-operating LArTPC in a neutrino beam producing a large dataset of GeV scale neutrino interactions.
The first important observation is that the calorimetric reconstruction performed by LArTPCs exhibits angular dependencies.
Since the charge is drifted in a specific direction, and projected onto wires oriented in three different directions, both the \dedx distribution that a perfect detector would measure and the precision and accuracy with which \dedx is actually measured depend on the track direction.
These effects are intrinsic to any wire-based readout LArTPC.
When not properly accounted for, they result in non-uniform and sub-optimal particle identification performance.
This work proposes a new likelihood-based method to perform particle identification which properly accounts for angular dependencies and mitigates their impact by effectively combining the calorimetric measurements performed on the three wire planes.
It does so through the calculation of a likelihood derived from the detailed \ub simulation.
The resulting PID score shows greater separation power between proton and muon-induced tracks, with smaller dependence on the track angle with respect to the previous PID method.
This is quantified as a 94\% proton selection efficiency with a 10\% muon mis-identification rate.
The novel PID method expands the physics reach of MicroBooNE, allowing highly effective separation of different final states, with only minor angular dependency, as demonstrated by selecting exclusive final states originated by different \numucc interaction modes.
Future \ub analyses will incorporate this method in the event selection strategy, leading to detailed cross section measurements.
This methodology can be exported to other present and future LArTPCs, making it an important ingredient to address the ambitious liquid argon neutrino physics program moving forward.

\acknowledgments
This research is supported by the U.S. Department of Energy, Office of Science, Office of High Energy Physics, under the Award Number DE-SC0007881.

This document was prepared by the MicroBooNE collaboration using the
resources of the Fermi National Accelerator Laboratory (Fermilab), a
U.S. Department of Energy, Office of Science, HEP User Facility.
Fermilab is managed by Fermi Research Alliance, LLC (FRA), acting
under Contract No. DE-AC02-07CH11359.  MicroBooNE is supported by the
following: the U.S. Department of Energy, Office of Science, Offices
of High Energy Physics and Nuclear Physics; the U.S. National Science
Foundation; the Swiss National Science Foundation; the Science and
Technology Facilities Council (STFC), part of the United Kingdom Research and Innovation;  and The Royal Society (United Kingdom).  Additional support for the laser calibration system and cosmic ray tagger was provided by the Albert Einstein Center for Fundamental Physics, Bern, Switzerland.
N.F. dedicates this paper to Federico Tonielli, a physicist, peer, and friend who inspired and taught many young physicists but disappeared too early for the contribution he could have made to science.

\clearpage
\bibliography{bib}








\end{document}